\pdfoutput=1
\documentclass[letterpaper]{JHEP3}
\usepackage[pdftex]{graphicx}
\usepackage{amsmath,amssymb}

\def\a{\alpha}
\def\b{\beta}
\def\g{\gamma} \def\G{\Gamma}
\def\d{\delta}
\def\e{\epsilon}
\def\la{\lambda} \def\La{\Lambda}
\def\th{\theta}
\def\s{\sigma}
\def\w{\omega} 

\def\adot{{\dot{\alpha}}}
\def\bdot{{\dot{\beta}}}

\def\abar{\bar{a}} \def\Abar{\bar{A}}
\def\ubar{\bar{u}}

\def\Dbar{\bar{D}}
\def\Wbar{\bar{W}}
\def\sbar{\bar{\sigma}}

\def\psibar{\bar{\psi}} 

\def\labar{\bar{\lambda}}
\def\thbar{\bar{\theta}}

\def\atilde{\tilde{a}} 

\def\A{\mathcal{A}}
\def\D{\mathcal{D}}
\def\F{\mathcal{F}}
\def\H{\mathcal{H}}
\def\K{\mathcal{K}}
\def\L{\mathcal{L}}
\def\M{\mathcal{M}}
\def\N{\mathcal{N}}
\def\O{\mathcal{O}}
\def\Nfour{\mathcal{N}\,{=}\,4}
\def\Ntwo{\mathcal{N}\,{=}\,2}
\def\None{\mathcal{N}\,{=}\,1}

\def\R{\mathbb{R}}
\def\Z{\mathbb{Z}}

\def\half{\tfrac{1}{2}}
\def\fourth{\tfrac{1}{4}}
\def\Nc{N_{\rm c}}

\def\dsl{\raise.15ex\hbox{/}\kern-.57em\partial}
\def\psl{\raise.15ex\hbox{/}\kern-.57em p}
\def\ksl{\raise.15ex\hbox{/}\kern-.57em k}

\def\vev#1{\langle#1\rangle}

\def\tr{\text{tr}}
\def\p{\partial}
\def\Im{\text{Im}}
\def\Re{\text{Re}}

\def\sumint#1#2{\sum\mspace{-25mu}{\int_{#1,#2}}}
\def\inlinesumint#1#2{\Sigma\mspace{-12.80mu}{\int_{#1,#2}}}

\def\doublet{\bigl(\begin{smallmatrix} \la \\ \psi \end{smallmatrix}\bigr)}

\title{
    Thermodynamics of {\boldmath $SU(2)$ $\Ntwo$} supersymmetric
    Yang-Mills theory
}

\author{
    Steve Paik\footnote{\tt paik@u.washington.edu}~ and 
    Laurence G. Yaffe\footnote{\tt yaffe@phys.washington.edu}\\
    Department of Physics, University of Washington, Seattle, WA 98195, USA
}

\abstract{
    The thermodynamics of four-dimensional $SU(2)$ $\Ntwo$ super-Yang-Mills theory
    is examined in both high and low temperature regimes.
    At low temperatures, compelling evidence is found 
    for two distinct equilibrium states related by a spontaneously 
    broken discrete {\it R}-symmetry.
    These equilibrium states exist because the
    quantum moduli space of the theory has two singular points where extra 
    massless states appear.
    At high temperature, a unique 
    {\it R}-symmetry-preserving equilibrium state is found.
    Discrepancies with previous results in the literature are explained.
}


\preprint{}

\begin{document}

\section{Introduction}

Certain quantum field theories are known to possess a continuous set of 
inequivalent ground states, or in other words, a quantum moduli space.
Every point in the moduli space corresponds to a vacuum state with
zero energy.
Turning on a non-zero temperature $T$ will generically ``lift'' moduli space,
leaving a much smaller set of thermal equilibrium states.
For some range of temperatures,
there may be a unique equilibrium state.
For other temperatures,
there may be multiple degenerate equilibrium states
related by spontaneously broken global symmetries.

One may define a thermal effective potential, or free energy functional,
using the same coordinates which parametrize the zero temperature
moduli space.
This will turn the flat $T=0$ zero energy surface into
a non-trivial $T>0$ free energy surface.
Equilibrium states correspond to the global minima of this
free energy surface.
Computing this free energy surface in an interacting theory is,
of course, non-trivial.
Supersymmetry provides little help,
since cancellations between bosonic
and fermionic particles in virtual processes are spoiled by their different 
statistics at non-zero temperature.
Nevertheless, two extreme limits are 
interesting and amenable to analytical calculation: 
arbitrarily low temperatures and asymptotically high temperatures. 
In the former case, one might expect the free energy surface to be a slight 
deformation away from the flat surface of moduli space.
What does this lift look like, and
where do the minima of the free energy surface lie?
In the high temperature regime, thermal fluctuations 
should have a disordering effect on the system.
Are spontaneously broken global symmetries restored?
At what temperature?

In this work, we examine the effects of thermal fluctuations on the 
equilibrium properties and realization of global symmetries
in the simplest asymptotically free supersymmetric gauge theory with a
continuous moduli space,
$SU(2)$ $\Ntwo$ supersymmetric Yang-Mills theory.
Much about this theory is known from the 
celebrated work of Seiberg and Witten \cite{SW1,SW2}.
The following features make it an attractive model for our purposes:
\begin{enumerate}\advance\itemsep -6pt
\item [(\textit{i})]
The quantum theory 
has a continuous moduli space of vacua --- it is a one-complex dimensional 
K\"ahler manifold parametrized by a single complex number, $u$. 
\item [(\textit{ii})]
Each vacuum describes a Coulomb phase
where the long distance dynamics is Abelian; there is no vacuum state
with long distance non-Abelian dynamics
({\it i.e.}, confinement).
\item [(\textit {iii})]
Generic ground states spontaneously break a discrete {\it R}-symmetry.
\item [(\textit {iv})]
Asymptotic freedom guarantees that vacua in the 
neighborhood of infinity on moduli space have weakly-coupled 
descriptions in terms of the light elementary fields. 
\item [(\textit {v})]
Two distinguished ``singular'' points in moduli space exist where
extra massless states with spin $\leq \half$ appear.
The corresponding particles are magnetically charged 
under the long-distance Abelian gauge group and may be interpreted
as magnetic monopoles or dyons.
For vacua in neighborhoods of these special points, a low energy 
effective description is strongly-coupled in terms of the elementary
fields.
However, a version of electric-magnetic duality provides a 
weakly-coupled formulation in terms of dual fields.
\end{enumerate}
The combination of asymptotic freedom and electric-magnetic duality
enables one to use weak coupling methods to explore the dynamics
both near and far from the singular points in moduli space.
There is a dynamically generated mass scale $\La$ in the theory
(analogous to $\La_{\rm QCD}$).
For temperatures much greater than $\La$, the free energy may be computed 
as an asymptotic expansion in the small effective gauge coupling $g^2(T)$.
For temperatures much less than $\La$, one may use appropriate
low energy effective descriptions near infinity,
or near the special points on moduli space, to compute the free 
energy as an expansion in the appropriate effective gauge coupling
(either $g^2(u)$ or its magnetic dual).

Thermal effects in $SU(2)$ $\Ntwo$ gauge theory at low temperature 
have previously been studied by Wirstam \cite{Wirstam}.
In this work,
it was asserted that
the free energy density was locally minimized asymptotically far out
on moduli space, and on circles of non-zero radius surrounding the
singular points.
It was not made clear which local minima represented the global minimum. 
The free energy surface found in Ref.~\cite{Wirstam}
is depicted on the left side of Figure~\ref{fig:uflow}.
In this figure, arrows depict directions of free energy decrease
({\em i.e.}, minus the gradient).
The picture implies non-monotonic behavior
as one moves from a singular point to infinity,
with a free energy barrier separating the large $u$ domain
from the region near the singular points,
and some sort of instability at the singular points.
Such features are unexpected and surprising.
One puzzle is why the free energy surface slopes downward to infinity.
Massive states get heavier as one moves further out on moduli space so,
in the absence of interactions, one would expect their contribution
to the pressure to decrease (since the associated particle density
falls exponentially due to Boltzmann suppression).
The free energy density is minus the pressure, so the decoupling
of massive states as one approaches the boundary of moduli space
should lead to a rising free energy.
Do interactions, in an asymptotically free theory,
really change this simple behavior?

A second puzzle concerns the circle of minima around each singular point.
What physical mechanism leads to this?
There is no continuous global symmetry whose action on moduli space
produces phase rotations around a singular point,
and whose spontaneous breaking could explain
such a circle of free-energy minima.

\FIGURE[t]{
    \label{fig:uflow}
    {\bf \hspace{0.01in} (a) results of Ref.~\cite{Wirstam} 
      \hspace{1.8in} (b) this work}
    \\[5pt]
    \centerline{\includegraphics[height=2.7in]{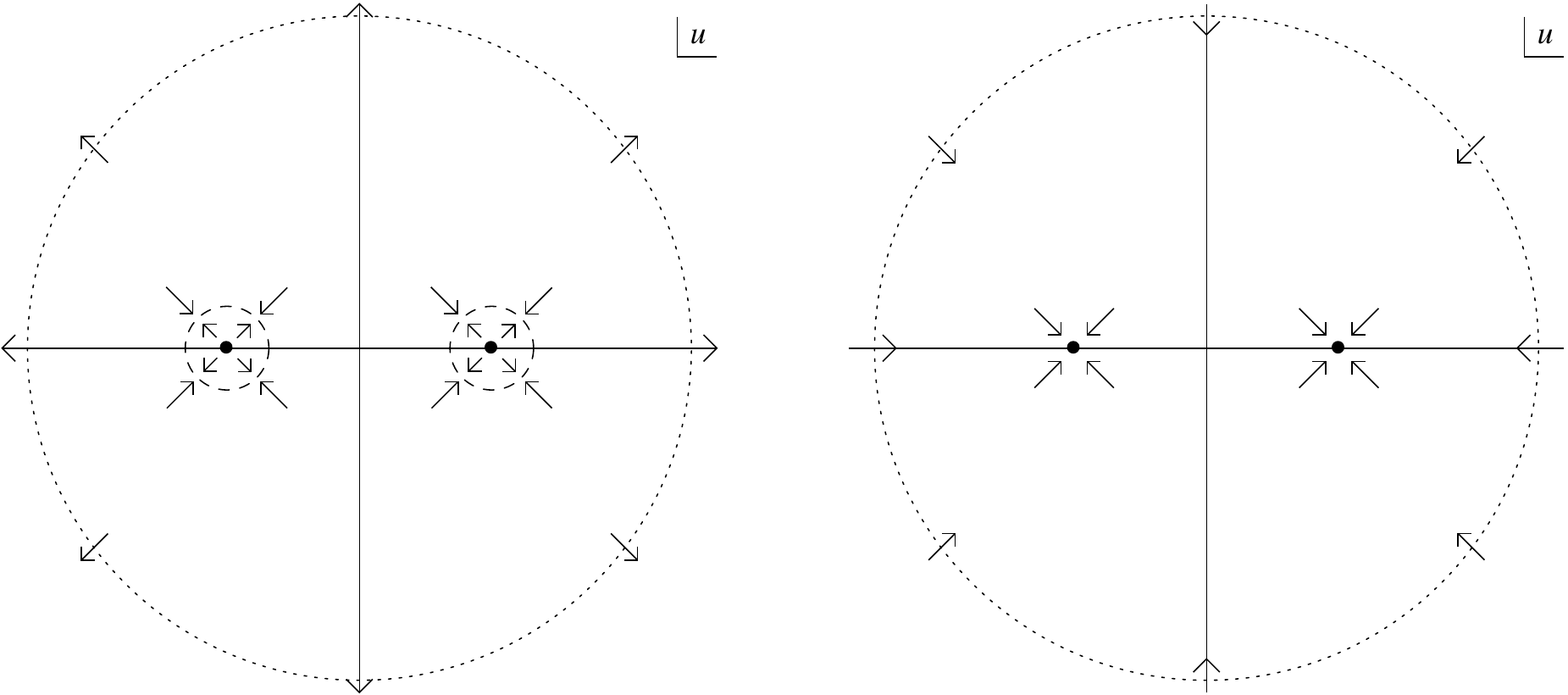}}
    \vspace*{-10pt}
    \caption{
    Qualitative form of the free energy surface in
    $SU(2)$ $\Ntwo$ Yang-Mills theory at $T \ll \La$.
    Arrows indicate directions on moduli space 
    for which free energy decreases. The singular points at $\pm u_0$ are
    represented by heavy dots. The large dotted circle at infinity serves to 
    guide the eye.
    Left: Asserted behavior from Ref.~\cite{Wirstam}. The dashed
    circles surrounding the singular points at $\pm u_0$ represent
    valleys of stable local minima. 
    Right: Results of our analysis.
    The singular points are stable minima.}
}

The main purpose of this paper is to derive the correct behavior
of the free energy surface at low temperatures by reconsidering
the computations of Ref.~\cite{Wirstam}.
Using effective field theory techniques, we systematically 
evaluate the contributions of successively longer wavelength fluctuations
to the effective scalar potential.
The free energy surface,
viewed as a functional of the translationally invariant expectation
value which parameterizes moduli space, may be
identified with this effective potential. 
Unlike Ref.~\cite{Wirstam}, we find the simple behavior
sketched on the right side of Figure \ref{fig:uflow}.
In the low temperature regime,
the asymptotic region of the free energy surface is locally unstable;
the free energy decreases as one moves inward from infinity.
The two singular points on moduli space where monopoles or dyons become
massless are local minima of the free energy.
There is no evidence for any other local minima.
Assuming so, this means that at sufficiently low temperatures there
are two distinct equilibrium states, related by a spontaneously broken
discrete \textit{R}-symmetry.
In contrast,
at sufficiently high temperatures
there is a unique equilibrium state and the 
{\it R}-symmetry is unbroken.
Hence, the theory undergoes a thermal phase transition.
The transition temperature must be a pure number times the
strong scale $\La$.

Our analysis mirrors that of Ref.~\cite{Wirstam},
but also extends it in several important 
areas in order to fix the misunderstanding of the free energy.
The problematic interpretation suggested by Figure \ref{fig:uflow}(a)
arises from
a sign error and a mistreatment of zero-frequency modes. 
We determine the sign of a 
crucial next-to-leading term in the momentum expansion of the low energy 
effective theory using general arguments based on 
analyticity constraints satisfied by scattering amplitudes in any UV-complete
theory \cite{Adams}. An $S$-duality transformation is then used to relate 
results in different regimes of moduli space \cite{Hen}.
When a large hierarchy separates the temperature from smaller momentum scales
of interest, we construct appropriate three-dimensional effective theories 
which implement the Wilsonian procedure of integrating
out short distance fluctuations in order to generate effective descriptions
for the long distance degrees of freedom \cite{BN1,BN2,ArnYaf,YamYaf}.

The remainder of this paper is organized as follows.
In Sec.~\ref{sec:review} we review relevant 
facts about $SU(2)$ $\Ntwo$ gauge theory at zero temperature,
and discuss the formulation of four-dimensional low energy
effective theories that will be useful for studying
the low temperature regime. In Sec.~\ref{sec:hightemp}, we discuss the high
temperature limit and the unique equilibrium state that realizes all 
{\it R}-symmetries.
Portions of this analysis involving the construction of the
appropriate high temperature three-dimensional effective theory
are relegated to Appendix~\ref{app:hightemp}.
Low temperature thermal effects on moduli space
are the subject of Sec.~\ref{sec:lowtemp}.
We analyze the thermal effective potential for
the scalar field that is related to the local 
coordinate on moduli space.
The generalization of our analysis to so-called $\Ntwo^*$ theory,
obtained by adding a single flavor of massive adjoint hypermultiplet
to $\Ntwo$ super-Yang-Mills,
is discussed in Sec.~\ref{sec:star}.
Finally, 
in Sec.~\ref{sec:discuss} we summarize our findings and discuss some open 
questions.

\section{Review of \boldmath $SU(2)$ $\Ntwo$ gauge theory}
\label{sec:review}

We consider four dimensional $\Ntwo$ supersymmetric pure Yang-Mills theory 
with gauge group $SU(2)$. It is renormalizable and asymptotically free.
Consequently, the dimensionless running coupling transmutates into a 
renormalization group invariant energy scale $\La$. This theory describes the 
interactions of an $\Ntwo$ vector multiplet $\A$. In terms of $\None$ 
superfields, the superfield $\A$ consists of a scalar-valued adjoint
representation chiral 
multiplet $\Phi$ and a spinor-valued chiral field strength $W_\a$. 
On-shell, the component fields in $\Phi$ are a complex 
adjoint scalar $\phi$ and an adjoint Weyl fermion $\psi_\a$.
The field strength $W_\a$ contains 
an adjoint Weyl fermion $\la_\a$ and a gauge field $A_\mu$.
We shall always work in Euclidean space unless noted otherwise, with an action%
\footnote{
    To obtain a Minkowski space Lagrange density, $\L_\text{(Mink.)}$, from the 
    Euclidean version, one performs the rotation $x^0 = -ix^0_\text{E}$ and 
    identifies $\L_\text{(Mink.)} = -\L_\text{(Eucl.)}$.
}
\begin{equation}
\label{action}
    S_\text{(Eucl.)} = \int dx^0_\text{E} \, d^3x \> \L_\text{(Eucl.)}.
\end{equation}
Using $\None$ superspace notation, the Lagrange density is given by
\begin{equation}
\label{Ls}
    -g^2\L = 
    \biggl(\int d^2\th\> \half\tr(W^\a W_\a) + \text{H.c.}\biggr) 
    + 2\int d^2\th \> d^2\thbar\> \tr(\Phi^\dag e^{[2V,\,\cdot\,]} \Phi)\,,
\end{equation}
where the field strength $W_\a$ and vector superfield $V$ are related by%
\footnote{
    We use the superspace conventions of Ref.~\cite{WB}. The factor of 2 
    in front of the K\"ahler potential ensures that the kinetic terms of the 
    two Weyl fermions have the same normalization.
}
\begin{equation}
    W_\a = -\tfrac{1}{8}\Dbar_\adot \Dbar^\adot (e^{-2V}D_\a e^{2V}) \,.
\end{equation}
We employ a matrix notation where all fields are Lie algebra-valued
(so, for example, $W_\a = W_\a^a T^a$ with repeated group indices summed over 
$a = 1,\dotsc,\dim G$). We take $G = SU(\Nc)$ and will specialize to $\Nc = 2$
momentarily. The fundamental representation Lie algebra generators $T^a$ are
traceless Hermitian $\Nc \times \Nc$ matrices satisfying
$[T^a, T^b] = if^{abc}T^c$, normalized such that $\tr(T^a T^b) = \d^{ab}/2$.
The structure constants are real and totally antisymmetric.
The integrands of the superspace integrals are manifestly 
invariant under gauge transformations of the form
$e^{2V} \to e^{-i\La}e^{2V}e^{i\La}$, where $\La$ is a fundamental 
representation chiral superfield. Under gauge transformations, the fields
$W_\a$, $\Phi$, and $e^{[2V,\cdot]}\Phi$ all transform via conjugation by 
the same group element.

Starting from Eq.~\eqref{Ls},
it is straightforward to show that the Lagrange density
in terms of on-shell component fields is given by
\begin{equation}
\label{Lc}
\begin{split}
    g^2\L 
    & = 2\, \tr\Bigl\{
    \fourth F_{\mu\nu}F_{\mu\nu} + i\labar\sbar_\text{E}^\mu D_\mu\la 
    + i\psibar\sbar_\text{E}^\mu D_\mu\psi + (D_\mu\phi)^\dag D_\mu\phi \\
    &\qquad
    - i\sqrt{2}[\la,\psi]\phi^\dag - i\sqrt{2}[\labar,\psibar]\phi
    + \half [\phi^\dag,\phi]^2 \Bigr\}\,.
\end{split}
\end{equation}
The covariant derivative $D_\mu = \p_\mu + i[A_\mu,\cdot]$, and
the field strength $F_{\mu\nu} = \p_\mu A_\nu - \p_\nu A_\mu 
+ i[A_\mu,A_\nu]$.
For a Weyl fermion, $\labar_\adot = (\la_\a)^\dag$ where
$\a,\adot = 1,2$. Our spinor conventions follow those of Ref.~\cite{WB} except
that the metric is $\d_{\mu\nu}$ and the matrix $\s_\text{E}^0$ is anti-Hermitian.
The matrices $(\s_\text{E}^\mu)_{\a\adot}$ form a basis for 
$2\times 2$ complex matrices: $\s_\text{E}^0 = i\bigl(\begin{smallmatrix}
-1 & \phantom-0 \\ \phantom-0 & -1 \end{smallmatrix}\bigr)$,
and $\s_\text{E}^i$ are the standard Pauli matrices.
The $\e$ tensor is used to raise spinor indices to obtain 
$(\sbar_\text{E}^\mu)^{\bdot\b} = \e^{\bdot\adot}\e^{\b\a}(\s_\text{E}^\mu)_{\a\adot}$. 
Numerically, $\sbar_\text{E}^0 = \s_\text{E}^0$ and $\sbar_\text{E}^i 
= -\s_\text{E}^i$, although the index structures are distinct.
Note that $[\la,\psi] = \la^\a\psi_\a - \psi^\a\la_\a$ and 
$[\labar,\psibar] = \labar_\adot\psibar^\adot - \psibar_\adot\labar^\adot$.

The Lagrange density \eqref{Lc} is invariant under a global $SU(2)_R \times 
U(1)_R$ {\it R}-symmetry. The fermions $\doublet$ transform as
a doublet under $SU(2)_R$ while $A_\mu$ and $\phi$ transform as singlets.%
\footnote{
    Since $\la$ and $\psi$ belong to different $\None$ multiplets, one may check
    the consistency of $\Ntwo$ supersymmetry in Eq.~\eqref{Lc} by testing the 
    invariance of the Lagrange density under
    $
    \doublet \to \bigl(\begin{smallmatrix} \phantom-\psi
			\\ -\la \end{smallmatrix}\bigr)
    $. 
    This discrete transformation corresponds to the
    $\bigl(\begin{smallmatrix} \phantom-0 & 1 \\ -1 & 0 \end{smallmatrix}\bigr)$ 
    element of $SU(2)_R$,
    which generates a $\Z_4$ subgroup.
} 
The $U(1)_R$ factor is an ordinary $\None$ {\it R}-symmetry
under which $\Phi$ has charge 2 and $W_\a$ has charge 1.
Quantum mechanically, the $U(1)_R$ is anomalous and only a 
$\Z_{4\Nc}$ subgroup survives.%
\footnote{
    This anomalous {\it R}-symmetry is the reason why no topological charge 
    density (and associated theta angle) appears in the Lagrange density.
    Since 
    $\Phi = \phi + \sqrt{2}\th\psi + \dotsb$, and $W_\a = -i\la_\a + \dotsb$,
    it follows that both Weyl fermions have {\it R}-charge 1. If one repackages
    these Weyl spinors as a single Dirac spinor
    $\Psi = \bigl(\begin{smallmatrix} \la_\a \\ \psibar^\adot \end{smallmatrix}
    \bigr)$, then $U(1)_R$ acts as a continuous chiral
    transformation, $\Psi \to e^{i\w\g_5}\Psi$, in a basis where $\g_5 = 
    \text{diag}(1,1,-1,-1)$. Under this field redefinition, the 
    fermion measure in the functional integral acquires a nontrivial 
    Jacobian involving the exponential of the topological charge. 
    Therefore, an appropriate choice of $\w$ allows one
    to cancel any dependence of the theory on the theta angle.
}
For $\Nc = 2$, the true global {\it R}-symmetry is thus 
$(SU(2)_R \times \Z_8)/\Z_2$,
where the division by $\Z_2$ is a reminder not to double count the 
$(-1)^F$ symmetry (with $F$ fermion number) 
present in both the center of $SU(2)_R$ and $\Z_8$.

It will be useful for later purposes to mention another massless representation
of $\Ntwo$ supersymmetry, the hypermultiplet
$\H$. In terms of $\None$ superfields, $\H$ consists of two scalar-valued chiral
multiplets $Q$ and $Q'$ that transform under conjugate representations of 
the gauge group. On-shell, $Q$ contains a complex scalar $q$ and a Weyl fermion
$\psi_q$. Similarly, $Q'$ contains a complex scalar $q'$ and a Weyl 
fermion $\psi_{q'}$. The scalars 
$\bigl(\begin{smallmatrix} q \\ q'^\dag \end{smallmatrix}\bigr)$ 
transform as an $SU(2)_R$ doublet while $\psi_q$ and $\psi_{q'}$ transform 
as singlets. Both $Q$ and $Q'$ have {\it R}-charge 0.

Vacua in this theory may be described classically by the requirements that 
$F_{\mu\nu} = \la = \psi = 0$, $\phi$ is covariantly constant, 
and $[\phi,\phi^\dag] = 0$. If a diagonalizing gauge transformation is made to
write $\phi = a \s^3/2$ for some arbitrary complex number $a$, then $\phi$
automatically commutes with its Hermitian conjugate. The two eigenvalues of
$\phi$ are $a$ and $-a$. Since we are free to permute them, $\phi = -a\s^3/2$ 
also describes the same vacuum. This permutation freedom is part of the residual gauge
invariance (specifically, conjugation by 
$\bigl(\begin{smallmatrix} \phantom- 0 & 1 \\ -1 & 0 \end{smallmatrix}\bigr)$).

A translation invariant and gauge invariant ``order parameter'' parameterizing the 
space of vacua is $u = \vev{\tr\,(\phi^2)}$. Since $u$ is a complex number, the
classical space of vacua is the complex $u$-plane. In the quantum theory 
$\phi$ (and its eigenvalues $\pm a$) is a fluctuating field.
The space of vacua may be changed by quantum effects, but it can 
never be entirely lifted. One reason is that it is impossible to generate an 
effective superpotential (and therefore no squares of F-terms in the scalar 
potential) invariant under $\Ntwo$ supersymmetry without also including at 
least one light hypermultiplet. There are no hypermultiplets at weak 
coupling. For $|\vev{a}| \gg \La$, asymptotic freedom ensures that the theory 
is weakly-coupled. When quantum fluctuations of the eigenvalues of $\phi$ are 
small compared to their vacuum expectation value, the fractional 
difference $\left<{(a-\vev a)^2}\right>/\vev{a}^2 \ll 1$ and hence
$u \approx \vev{a}^2/2$.

For any vacuum satisfying $|u| \gg \La^2$, a weak-coupling mean 
field analysis is reliable and shows that the Higgs mechanism 
reduces the gauge group from $SU(2)$ to $U(1)$.
There are massive $W$ bosons charged under the $U(1)$ photon.
The $W$ bosons have 
masses proportional to the expectation value of $a$,
\begin{equation}
\label{Wmass}
    M_W = \sqrt{2}\,|\vev{a}|\,.
\end{equation}
$\Ntwo$ supersymmetry requires the $W$ bosons
to belong to Abelian vector multiplets, and other components in the multiplet 
must have the same mass. The dynamics of the resulting theory at
momenta much less than $M_W$ is both Abelian and $\Ntwo$ supersymmetric. The 
$\Z_8$ {\it R}-symmetry is spontaneously broken since $u$ has {\it R}-charge 4.
The unbroken subgroup $\Z_4$ acts trivially on $u$,
while the coset $e^{2\pi i/8} \Z_4$ acts as $u \to -u$.

In Ref.~\cite{SW1}, it was shown that when quantum effects are taken into account 
the space of vacua, or moduli space, is precisely the $u$-plane but with three 
singular points: one at infinity and two at finite values $\pm u_0$. 
One may choose to renormalize the operator $\tr\,(\phi^2)$ so that $u_0 = \La^2$.
The existence of a continuous set of vacua implies that $\vev{\tr\,(\phi^2(x))}$
may have arbitrarily long wavelength fluctuations. Such configurations
can have arbitrarily small spatial gradients with negligible cost in energy,
implying that there are massless states in the spectrum of the Hamiltonian. These states
comprise an $\Ntwo$ Abelian vector multiplet. The singularity at infinity is a 
consequence of asymptotic freedom. The singularities at $\pm u_0$ are 
interpreted as vacua in which extra massless states appear in the spectrum. 
These massless states have spins 0 and $\half$, and constitute an $\Ntwo$ 
Abelian hypermultiplet. Since
there are no elementary hypermultiplets in the theory, these extra particles 
are solitonic excitations. Massless non-Abelian gluons never appear for any
choice of $u$. That is, there is no vacuum corresponding to an infrared fixed 
point with conformal invariance. Every choice of $u$ (even $u = 0$) corresponds
to a theory in which the long distance dynamics is Abelian, possibly with 
extra massless excitations. The theory is always in a Coulomb phase.

To discuss the particle spectrum at a given $u$, it is helpful to 
construct an effective theory describing the dynamics in such a vacuum
at arbitrarily low momentum. At a generic point in moduli space, the 
massless fields comprise a $U(1)$ $\Ntwo$ vector 
multiplet which is simply the neutral component $\A^3 = (\Phi^3, W^3_\a)$ of 
the gauge triplet.%
\footnote{
    This is a direct consequence of the Higgs mechanism for large $|u|$.
    By analytic continuation in the $u$-plane (avoiding possible singularities
    or cuts), it must also be true even for $|u| \sim \La^2$ where the dynamics
    is strongly-coupled. In fact, if this were not true, then the $U(1)$ 
    photon would have to obtain a mass through some type of Higgs mechanism.
    This cannot happen because there are no charged scalars in the $U(1)$ 
    vector multiplet~\cite{Argy}.
}
The complex scalar field $\phi^3 = \Phi^3|_{\th=\thbar=0}$ 
is identical to the eigenvalue field $a$ when $\phi$ is diagonal. Abusing
notation (but following Ref.~\cite{SW1}),
we shall henceforth refer to $\A^3$ as $\A$, its $\None$ 
scalar-valued chiral multiplet $\Phi^3$ as $A$, and its field strength 
$W^3_\a$ as $W_\a$. The low energy effective theory
possesses $\Ntwo$ supersymmetry, and this is made manifest by constructing 
a Lagrange density directly in $\Ntwo$ superspace,
\begin{equation}
\label{Leffss}
    \L_\text{eff} = -\tfrac{1}{4\pi}\>\Im\int d^4\th\, \F(\A) 
    - \int d^4\th \> d^4\thbar\> \K(\A,\bar{\A}) + O(n \geq 6).
\end{equation}
The prepotential $\F(\A)$ is a mass 
dimension two holomorphic function of $\A$.
The function $\K(\A,\bar{\A})$ is 
dimensionless and non-holomorphic in $\A$.%
\footnote{
    Henceforth, $\bar{\A} \equiv \A^\dag$.
}
The terms in Eq.~\eqref{Leffss} are organized as an expansion in the 
`order in derivatives' $n$, explained in Ref.~\cite{Hen}. The number $n$ is defined 
such that $\A$ has $n = 0$ and the supercovariant derivative has $n = 1/2$. 
Ordinary spacetime derivatives have $n = 1$ since they arise from 
anticommutators of supercovariant derivatives. From the structure of a 
supercovariant derivative, it immediately follows that Grassmann-valued 
superspace coordinates $\th^\a_i, \thbar^{\adot i}$ have $n = -1/2$. 
Gauge fields and scalars have $n=0$ and fermions have $n=1/2$.
Based on this counting scheme, the chiral superspace integral has $n = 2$ 
and the full superspace integral has $n = 4$. Therefore, knowledge of the 
prepotential completely determines terms in the effective Lagrange
density with up to two spacetime derivatives and at most four fermions. 
Note that Eq.~\eqref{Leffss} is unchanged by a linear shift of $\F$ or the 
addition of a holomorphic function (and its Hermitian conjugate) to $\K$.
When $\F$ and $\K$ are determined through matching calculations at the 
momentum scale $M_W$, then Eq.~\eqref{Leffss} will correctly reproduce gauge 
invariant correlators of the light fields for distances $\gg M_W^{-1}$.

The effective Lagrange density \eqref{Leffss} in $\None$ superspace notation
is
\cite{SW1,Hen}
\begin{equation}
\label{Leffs}
    \L_\text{eff} = \L_\text{eff}^{n=2} + \L_\text{eff}^{n=4} + O(n\geq 6),
\end{equation}
where
\begin{equation}
\label{Leff2s}
    \L_\text{eff}^{n=2} = 
    -\tfrac{1}{4\pi}\Im\biggl[\int d^2\th\,\half\F''(A)\, W^\a W_\a + 
    \int d^2\th d^2\thbar\, \F'(A)\, \Abar\biggr],
\end{equation}
and
\begin{equation}
\label{Leff4s}
\begin{split}
    \L_\text{eff}^{n=4} =& 
    -\int d^2\th \> d^2\thbar\,\Bigl\{
    \K_{A\Abar}(A,\Abar)\Bigl[(D^\a D_\a A) (\Dbar_\adot \Dbar^\adot\Abar)
    + 2(\Dbar_\adot D^\a A) (D_\a \Dbar^\adot \Abar) \\
    & ~~~~~~~~~~~~~~~~~~~~~~~~~~~~~~~~~
    + 4(D^\a W_\a) (\Dbar_\adot \Wbar^\adot) \\
    & ~~~~~~~~~~~~~~~~~~~~~~~~~~~~~~~~~
    - 4(D^{(\a}W^{\b)}) (D_{(\a}W_{\b)}) - 2D^\a D_\a(W^\b W_\b) \\
    & ~~~~~~~~~~~~~~~~~~~~~~~~~~~~~~~~~
    - 4(\Dbar_{(\adot}\Wbar_{\bdot)}) (\Dbar^{(\adot}\Wbar^{\bdot)})
    - 2\Dbar_\adot\Dbar^\adot(\Wbar_\bdot\Wbar^\bdot)\Bigr] \\
    & ~~~~~~~~~~~~~~~~~~
    - 2\K_{AA\Abar}(A,\Abar) W^\a W_\a D^\b D_\b A
    - 2\K_{A\Abar\Abar}(A,\Abar) \Wbar_\adot\Wbar^\adot 
    \Dbar_\bdot \Dbar^\bdot \Abar \\
    & ~~~~~~~~~~~~~~~~~~
    + \K_{AA\Abar\Abar}(A,\Abar)\Bigl[-8(W^\a D_\a A)
    (\Wbar_\adot\Dbar^\adot\Abar) + 4W^\a W_\a \Wbar_\adot \Wbar^\adot\Bigr]
    \Bigr\}.
\end{split}
\end{equation}
In expression \eqref{Leff4s}, subscripts on the non-holomorphic function $\K$ 
indicate derivatives with respect to the indicated arguments. 
Expression \eqref{Leff4s} is unique up to terms proportional
to $D^\a W_\a - \Dbar_\adot\Wbar^\adot$ which vanish because $W_\a$ satisfies 
the Bianchi identity. 

A gauge invariant description of moduli space is the $u$-plane, a one-complex 
dimensional manifold. One may promote the coordinate $u$ to a field $u(x)$, 
valued in the complex numbers. The dynamics of arbitrarily 
long wavelength fluctuations of $u(x)$ around a constant value is described by 
a sigma model action of the form $S_\text{eff} = \int d^4x\bigl[\g(u,\bar{u})
\, \p_\nu u \, \p_\nu\bar{u} + \dotsb\bigr]$. One can regard the coefficient of the
two derivative term as a metric on moduli space. The line element is written as
$ds^2 = \g(u,\bar{u}) \, du \, d\bar{u}$. 

The metric on moduli space is easily determined for $|u| \gg \La^2$. 
In this regime the effective theory is formulated in terms of the scalar 
field $a(x)$, and its interactions are weakly coupled due to asymptotic 
freedom. The translationally invariant vacuum expectation value $\vev{a}$ 
is mapped to $u$ by the approximate formula $u \approx \vev{a}^2/2$. Therefore,
the asymptotic region of moduli space may be parametrized by the local 
coordinate $a$.%
\footnote{
    Here $a$ is understood to mean $\vev{a}$. Future usage should be clear from
    context.
}
The induced metric in field configuration space is obtained from the K\"ahler 
potential $K(A,\Abar) = \frac{1}{4\pi}\Im(\F'(A)\Abar)$. The full superspace 
integral of the K\"ahler potential yields
\begin{equation}
\L_\text{eff} = \g(a,\abar)\, \p_\mu a \, \p_\mu\abar + \dotsb, 
\end{equation}
with the K\"ahler metric
\begin{equation}
\g(a,\abar) = \p_a \p_{\abar} K|_{\th=\thbar=0} = \tfrac{1}{4\pi} \Im\,\F''(a)\,.
\end{equation}
However, the effective theory is more than just a 
sigma model; it is also an Abelian gauge theory. Define a holomorphic gauge 
coupling function $\tau(A) = \F''(A)$. The half superspace integral of 
$\frac{1}{8\pi}\tau(A)\,W^\a W_\a$ yields 
\begin{equation}
\L_\text{eff} = \fourth g_\text{eff}^{-2}(a,\abar)
F_{\mu\nu}F_{\mu\nu} + \tfrac{1}{32\pi^2}\,\th_\text{eff}(a,\abar)
F_{\mu\nu}\widetilde{F}_{\mu\nu} + \dotsb,
\end{equation}
where the inverse effective gauge coupling and the
effective theta angle are given by
\begin{equation}
g_\text{eff}^{-2}(a,\abar) = \tfrac{1}{4\pi}\, \Im\,\tau(a), \qquad
\th_\text{eff}(a,\abar) = 2\pi\,\Re\,\tau(a).
\end{equation}
Note that $\Ntwo$ supersymmetry requires that the inverse gauge
coupling $g_\text{eff}^{-2}$ and the K\"ahler metric $\g$ coincide.

The prepotential fixes all the effective couplings in $\L_\text{eff}^{n=2}$. 
It is straightforward to express expression \eqref{Leff2s} in terms of 
off-shell component fields,
\begin{equation}
\label{Leff2c}
\begin{split}
    \L_\text{eff}^{n=2} & =  
    g_\text{eff}^{-2}(a,\abar)\Bigl\{
    |\p_\mu a|^2 + \fourth F_{\mu\nu}^2 + 
    \bigl(\tfrac{i}{2}\psi\s_\text{E}^\mu D_\mu\psibar 
    + \tfrac{i}{2}\la\s_\text{E}^\mu D_\mu\labar + \text{H.c}\bigr) 
    - |F|^2 - \half D^2 \\
    &\quad 
    + \bigl[\G(a,\abar)\bigl(\half\psi^2 F^* + \half \la^2 F 
    - \tfrac{i}{\sqrt{2}}\la\psi D 
    - \tfrac{1}{\sqrt{2}}\la\s_\text{E}^{\mu\nu}\psi F_{\mu\nu}\bigr)+\text{H.c.}\bigr] 
    \\
    &\quad
    - \bigl[R(a,\abar)\fourth\la^2\psi^2 + \text{H.c}\bigr]\Bigr\}
    + \tfrac{1}{32\pi^2}\,\th_\text{eff}(a,\abar)\,
    F_{\mu\nu}\widetilde{F}_{\mu\nu}.
\end{split}
\end{equation}
In expression \eqref{Leff2c}, each fermion bilinear is shorthand for a spinor
contraction 
({\it e.g.}, $\psi^2 \equiv \psi^\a\psi_\a$), the Abelian field strength
$F_{\mu\nu} \equiv \p_\mu A_\nu - \p_\nu A_\mu$ and its Hodge dual
$\widetilde{F}^{\mu\nu} \equiv \half\e^{\mu\nu\a\b}F_{\a\b}$
(with $\e^{0123} \equiv 1$), and
\begin{subequations}
\begin{align}
D_\mu &= \p_\mu - \G(a,\abar)\, \p_\mu a \,, \\
\G(a,\abar) &= \g(a,\abar)^{-1}\, \p_a \g(a,\abar) \,, \\ \nonumber
R(a,\abar) &= \p_a \G(a,\abar) + \G(a,\abar)^2 \\
&= \g(a,\abar)^{-1}\, \p_a^2 \g(a,\abar)\,.
\end{align}
\end{subequations}
The on-shell form of $\L_\text{eff}^{n=2}$ may be obtained by solving the 
equations of motion for the auxiliary fields, but we will not need that
result.%
\footnote{
    The D-term equation is 
    $D = -\frac{1}{\sqrt{2}}\G(a,\abar)i\la\psi + \text{H.c.}$
    and the F-term equation is $F = \half\G(a,\abar)\psi^2 + \half\G(a,\abar)^*
    \labar^2$. Substituting these into expression \eqref{Leff2c} produces 
    four-fermion operators only.
    We do not need the explicit result since we will
    carry out perturbative calculations using Feynman rules
    for off-shell fields.
} 

The prepotential may be determined as follows. The one-loop beta function for 
the running gauge coupling of $SU(2)$ $\Ntwo$ gauge theory
is
\begin{equation}
\label{betafn1}
    \mu \frac{dg^2}{d\mu} = -\frac{1}{2\pi^2} \, g^4\Bigl[1 + O(e^{-8\pi^2/g^2})\Bigr].
\end{equation}
Higher order loop corrections vanish due to supersymmetry, but we have included
the form of nonperturbative one-instanton corrections. Integrating 
Eq.~\eqref{betafn1} yields
\begin{equation}
g^{-2}(\mu) = \frac{1}{4\pi^2}\ln(\mu^2/\La^2) 
+ \text{const.} + O(\La^4/\mu^4) \,,
\end{equation}
where $\La$ is the conventional definition
of the strong scale. This may be matched to the effective gauge 
coupling of the low energy effective theory at the $W$ mass scale. 
That is, at $\mu = M_W$ one has $g_\text{eff}^2(a,\abar) = g^2(M_W)$. For 
asymptotically large $|u|$, the mass formula Eq.~\eqref{Wmass} implies
\begin{equation}
    \Im\,\tau(a) \approx \frac{1}{\pi}\ln\biggl(\frac{|a|^2}{\La^2}\biggr).
\end{equation}
This leading log is reproduced by a prepotential
\begin{equation}
\label{prepot}
    \F(a) \approx \frac{i}{2\pi}\, a^2\ln\biggl(\frac{a^2}{\La^2}\biggr).
\end{equation}

The line element on moduli space is $ds^2 = \frac{1}{4\pi}\Im\,\tau(a) 
da d\abar$, where the metric is given explicitly by $\frac{1}{4\pi}\Im\,\tau(a)
\approx \frac{1}{4\pi^2}[\ln(|a|^2/\La^2) + 3]$. The metric is single-valued 
and positive for $|a| \gg \La$. It diverges as $|a|/\La \to \infty$ which 
means that the effective gauge coupling becomes arbitrarily small. This is just
a restatement of asymptotic freedom. For smaller values of $|a|$ there is a 
difficulty: the metric is negative. In fact, $\tau(a)$ is 
holomorphic so $\Im\,\tau(a)$ must be harmonic, and since it is not constant,
it must be unbounded from below. The metric fails to be 
positive-definite, or equivalently, $g_\text{eff}$ fails to be real.
One is forced to concede that $a$ is valid as a local coordinate only 
asymptotically far out on moduli space. A key 
observation of Ref.~\cite{SW1} is that the form of the metric, as well as its 
positivity, can be maintained if an additional coordinate on moduli space 
is introduced: $a_D = \p\F(a)/\p a$. Then the line element on moduli
space may be expressed as
$ds^2 = \frac{1}{2i}(da_D \, d\abar - da \, d\abar_D)$ 
which exhibits a symmetry under 
$\bigl(\begin{smallmatrix} a_D \\ a \end{smallmatrix}\bigr) \to 
\bigl(\begin{smallmatrix} a \\ -a_D \end{smallmatrix}\bigr)$. This allows one
to also use $a_D$ as a local coordinate on moduli space, with a different 
harmonic function serving as the metric. The region of the $u$-plane
in which $a_D$ is a good coordinate will be discussed shortly.

The metric on moduli space is preserved by real linear fractional
transformations of the coordinates $\bigl(\begin{smallmatrix} a_D \\ a 
\end{smallmatrix}\bigr)$. Define a holomorphic vector field 
$\vec{a}(u) = \bigl(\begin{smallmatrix} a_D(u) \\ a(u) \end{smallmatrix}\bigr)$
over the $u$-plane. The induced metric is given by
$ds^2 = -\frac{i}{2}\, \e_{mn} \frac{da^m}{du}\frac{d\abar^n}{d\ubar} \,
du \, d\ubar$,
and is preserved by monodromies $M \in SL(2,\Z)$ that act on the 
vector field as $\vec{a} \to M\vec{a}$.%
\footnote{
    The invariance of the mass formula for BPS-saturated states under the 
    action of monodromies implies that 
    $M$ must be integer-valued, and that no constant vector can be added
    to the linear transformation $\vec{a}\to M\vec{a}$.
    The actual monodromy group 
    turns out to be $\G(2) \subset SL(2,\Z)$ which consists of matrices 
    congruent to the identity matrix (modulo 2, taken element-wise) \cite{SW1}.
}

The presence of monodromy and $\Ntwo$ supersymmetry naturally imply a type of
electric-magnetic duality. This fact is uncovered by asking the following 
question: if monodromies rotate $a$ into $a_D$, but $a$ belongs to a vector 
multiplet, then what effect does an $SL(2,\Z)$ transformation have on $A_\mu$? 
In particular, consider the monodromy
\begin{equation}
S = \begin{pmatrix} \phantom-0 & 1 \\ -1 & 0 \end{pmatrix}
\end{equation}
which
fully rotates $a$ into $-a_D$. In Ref.~\cite{SW1},
it was demonstrated that $S$ acts 
on the gauge fields as a Fourier transformation in field configuration space 
from $A_\mu$ to a dual gauge field $A_{D\mu}$. The form of the action remains
the same in the new variables except that the effective gauge coupling inverts
(as a consequence of integrating a Gaussian functional). In summary, $SL(2,\Z)$
acts linearly on the $\None$ chiral multiplets $A$ and $A_D = \p\F(A)/\p A$, and
by electric-magnetic duality on the $\None$ chiral field strengths $W_\a$ and 
$W_{D\a}$. Just as $\L_\text{eff}^{n=2}$ contains two derivative terms for 
the effective theory asymptotically far out on moduli space, one can define 
an analogous Lagrange density for the $S$-dual effective theory,
\begin{equation}
    \L_{D,\,\text{eff}}^{n=2} = -\tfrac{1}{4\pi}\,\Im\biggl[
    \int d^2\th\> \half\,\F_D''(A_D) W^\a_D W_{D\a}
    + \int d^2\th\, d^2\thbar\> \F_D'(A_D)\Abar_D\biggr].
\end{equation}
This description is useful precisely where the original gauge coupling blows
up. Thus, the $S$-dual effective theory is valid near the massless monopole
point in moduli space. The dual prepotential $\F_D$ may be related to the 
original prepotential via a Legendre transform \cite{Hen}. In practice,
one may obtain $\F_D(A_D)$ using a method similar to the one used to obtain 
$\F(A)$.

Four derivative terms in the $S$-dual effective theory are given by the
Lagrange density $\L_{D,\,\text{eff}}^{n=4}$. The explicit form for 
$\L_{D,\,\text{eff}}^{n=4}$ is identical in structure to expression 
\eqref{Leff4s} except
that dual chiral superfields $A_D$ appear in place of elementary ones, and the
dynamics is determined by a non-holomorphic function $\K_D$. A
priori, there is no relation between $\K$ and $\K_D$. However, in 
Ref.~\cite{Hen} it was proved directly in $\Ntwo$ superspace that the 
non-holomorphic function $\K$ is a modular function with respect to $SL(2,\Z)$.
In particular, under an $S$ transformation (which amounts to a Fourier 
transformation),
\begin{equation}
\label{SonK}
    \K_D(-A_D, -\Abar_D) \equiv \K(A, \Abar)\,.
\end{equation}
This relation will be important later.

The $\Ntwo$ theory has a BPS bound, $M \geq \sqrt{2}|Z|$, where $Z$ is the 
central charge in the extended supersymmetry algebra. The lightest states 
saturate this bound. A state with electric charge $n_e$ (defined by its
coupling to the photon $A_\mu$) and magnetic charge $n_m$ has mass
\begin{equation}
\label{BPSmass}
    M = \sqrt{2}\,|Z|,
\end{equation}
where $Z(u) = a(u)\, n_e + a_D(u)\, n_m$. Since $Z$ determines particle masses, 
it is renormalization group invariant.

The singularities in the finite $u$-plane arise from massive $\Ntwo$ Abelian 
hypermultiplets that become exactly massless at $u = \pm u_0$ \cite{SW1}. 
At $u_0$, a magnetic monopole with charges $(n_m,n_e) = (1,0)$ becomes 
massless. At $-u_0$, a dyon with charges $(n_m,n_e) = (1,-1)$ becomes 
massless.%
\footnote{
    In Ref.~\cite{SW1} these charge assignments for the extra massless multiplets 
    were shown to pass several consistency checks. For
    instance, there is a monodromy around each singularity in the $u$-plane
    and the set of monodromies should furnish a representation of the 
    fundamental group of the $u$-plane (with singularities deleted) in 
    $SL(2,\Z)$. If the monodromies around $u_0$ and $-u_0$ are calculated 
    assuming that the hypermultiplets becoming massless are a $(1,0)$ monopole 
    and $(1,-1)$ dyon, respectively, then the monodromies indeed generate a 
    subgroup of $SL(2,\Z)$. As another check, the triangle inequality and 
    conservation of energy ensure that when the ratio $a_D/a$ is not real, 
    the monopole and dyon cannot decay because $n_m$ and $n_e$ are 
    relatively prime. Asymptotically far out on moduli space, 
    there exist field configurations with magnetic charge --- these 
    are the semiclassical monopoles. Moving 
    in from infinity along the real $u$-axis toward $\pm u_0$, one never 
    crosses a curve on which $a_D/a$ becomes real. This implies that whatever
    stable BPS-saturated states exist at infinity must also appear in the
    strongly-coupled region.}
Let us study the monopole. According to Eq.~\eqref{BPSmass}, the 
monopole has a mass proportional to the coordinate $a_D$, 
\begin{equation}
\label{Mmass}
    M_\text{m} = \sqrt{2}\,|a_D|.
\end{equation} 
This mass vanishes at $u_0$ only if $a_D(u_0) = 0$. In the vicinity of the 
point $u_0$, the low energy effective theory must describe 
a $U(1)$ $\Ntwo$ gauge theory coupled to a massive hypermultiplet. 
This theory is essentially an $\Ntwo$ 
version of QED with the light monopoles playing the role of ``electrons.'' It
is infrared free and the renormalization of the dual effective gauge coupling 
is due primarily to one-loop photon self-energy diagrams where the fields of 
the light hypermultiplet run around the loop. Consequently, the renormalization
group implies that each decade of momentum, from a fixed UV cutoff down to 
the monopole mass, contributes the same amount to the inverse coupling.
Since the monopole mass vanishes as $u \to u_0$, it follows that the dual
effective gauge coupling $g_D$ vanishes as $u \to u_0$. Hence, $a_D$ is a good 
coordinate on moduli space in a neighborhood of the point $u_0$.

To determine the dual prepotential, consider the beta function of $\Ntwo$ QED 
with a single hypermultiplet,
\begin{equation}
\label{betafn2}
    \mu\frac{dg^2_D}{d\mu} = \frac{1}{4\pi^2}\,g_D^4.
\end{equation}
Integrating Eq.~\eqref{betafn2} yields $g_D^{-2}(M_\text{m}) = 
\frac{1}{8\pi^2}\ln(\La^2/M_\text{m}^2) + \text{const.}$ 
This result holds for $u$ close to $u_0$. The inverse gauge 
coupling is related to the imaginary part of a holomorphic function
$\tau_D(A_D) = \F_D''(A_D)$. Plugging in Eq.~\eqref{Mmass} yields
\begin{equation}
    \Im\,\tau_D(a_D) \approx 
    -\frac{1}{2\pi}\ln\biggl(\frac{|a_D|^2}{\La^2}\biggr).
\end{equation}
This leading log is reproduced by a dual prepotential
\begin{equation}
    \F_D(a_D) \approx -\frac{i}{4\pi}\,a_D^2\ln\biggl(\frac{a_D^2}{\La^2}\biggr).
\end{equation}

An explicit expression for the four derivative terms in $\L_\text{eff}^{n=4}$,
including a formula for the non-holomorphic function $\K$, is discussed in 
Sec.~\ref{sec:lowtemp}.

Because there are $R$-symmetry transformations which act
on the entire vacuum manifold as
$u \to -u$, the behavior of the free energy density near the
massless monopole and dyon points is identical. Consequently, it is unnecessary 
to write down a low energy effective theory incorporating light dyons. Our low 
temperature analysis 
relies only on the weakly-coupled effective descriptions near $u = \infty$ and 
$u = u_0$.

\section{High temperature behavior}
\label{sec:hightemp}

$SU(2)$ $\Ntwo$ gauge theory at asymptotically high temperatures,
\begin{equation}
\label{hierarchy}
T \gg gT \gg g^2T \gg \La\,,
\end{equation}
is weakly coupled, $g \ll 1$, on length scales small compared to $1/(g^2T)$.
[Here and henceforth, $g \equiv g(T)$ stands for the running gauge coupling 
evaluated at the scale $T$.]
Given the hierarchy of scales \eqref{hierarchy}, one may compute
the dependence of the free energy density $F/V$
on the (translationally invariant) thermal expectation value of the 
complex scalar field $\phi$,
using effective field theory techniques and perturbation theory.
In other words, one may compute the thermal effective potential
for $\phi$.
Minimizing $F/V$ with respect 
to $\vev{\phi}$ determines the number and location of equilibrium states,
which in turn determine
the realization of the discrete {\it R}-symmetry.

High temperature perturbation theory, at $\vev\phi = 0$,
shows that the non-Abelian $\Ntwo$ plasma has a positive Debye mass,
\begin{equation}
\label{staticmass}
    m_\text{D}^2 = 2g^2T^2 + O(g^4T^2) \,,
\end{equation}
and that the field $\phi$ also develops an effective thermal mass,
\begin{equation}
\label{scalarmass}
    m_\phi^2 = g^2T^2 + O(g^4T^2)\,.
\end{equation}

Since the curvature of the thermal effective potential for $\phi$,
at $\vev\phi = 0$, equals the thermal mass $m_\phi^2$,
the positive value \eqref{scalarmass} indicates that $\phi = 0$
is a local minimum of the free energy.
To demonstrate that this is, in fact, the global minimum,
one must evaluate the effective potential for $\phi$ arbitrarily far
away from $\phi = 0$.
This we do in Appendix \ref{app:hightemp}.
The result is unsurprising:
for asymptotically high temperatures
there is a unique equilibrium state at $\vev{\phi} = 0$.
The free energy density in this equilibrium state has the asymptotic expansion
\begin{equation}
\label{FEhigh3}
\begin{split}
    F/V &= 
%%%    (T/g_3^2)\, f + 3\left[I(m_\text{E}^2)\,T + 2I(m_\phi^2)\,T\right]
%%%\\ &=
    3T^4\biggl[-\frac{\pi^2}{12} + \frac{g^2}{8}
    -\frac{1 + \sqrt{2}}{6\pi}\, g^3 + O(g^4)\biggr].
\end{split}
\end{equation}
The leading term is the ideal gas blackbody contribution.
The $O(g^2)$ term comes from two-loop contributions
at the momentum scale of $T$, while the $O(g^3)$ term
arises from zero frequency contributions on the scale of $gT$.

The positive thermal mass (squared) \eqref{scalarmass} and
the unique minimum of the thermal effective potential
imply that the discrete {\it R}-symmetry,
which is spontaneously broken at zero temperature,
is restored for sufficiently high temperature,
$T \gg \La$.%
\footnote
    {%
    A similar conclusion was reached in Ref.~\cite{Pastras}.
    However, this analysis did not include the effects of interactions.
    }
To argue this formally,
recall that the gauge invariant order parameter involves the operator
$\tr\,(\phi^2)$.
To probe spontaneous symmetry breaking one may consider 
the correlator $D(\vec{x}{-}\vec{y}) \equiv
\vev{\tr\,(\phi(\vec{x})^\dag)^2\> \tr\,(\phi(\vec{y})^2)}$,
where $\vev{\dotso}$ represents an expectation in a $R$-symmetry invariant 
thermal equilibrium state (which might be a statistical mixture
of two noninvariant pure states).
A non-vanishing large distance limit,
$
    D(\vec{x}{-}\vec{y}) \mathop{\rlap{\:\:/}{\longrightarrow}} 0
$
as $|\vec{x}{-}\vec{y}|\to\infty$,
would indicate breakdown of cluster decomposition and consequent
spontaneous symmetry breaking.
However, the positive thermal mass \eqref{scalarmass}
implies that scalar correlators, evaluated at the
$\vev{\phi} = 0$ global minimum of the effective potential,
fall exponentially fast at distances large compared to $m_\phi^{-1}$. 
Consequently, $D(\vec{x}-\vec{y})$ approaches zero 
at large distances and does not contain a disconnected part,
signaling unbroken $R$ symmetry.

\section{Low temperature effective theory}
\label{sec:lowtemp}

The spectrum and low energy dynamics of $SU(2)$ $\Ntwo$ gauge theory varies 
drastically over moduli space. Far out on moduli space, near the asymptotically
free vacuum, charged particle masses become arbitrarily large
and the low energy dynamics contains only an Abelian massless vector multiplet. 
Near the strongly-coupled vacua there are, in addition to the 
massless vector multiplet, charged hypermultiplets consisting of light
magnetic monopoles or dyons. Whether or not this extra matter affects the 
low energy dynamics is determined by the ratio of the monopole (or dyon)
mass to the temperature.
We will focus attention on
three distinct regions of the free energy surface where weakly-coupled
effective theories may be constructed.
Using these effective theories, our goal is to
compute the free energy density as a 
function of the moduli space coordinate $u$.

\subsection{Near the singularity at infinity}
\label{subsec:infty}

Consider the asymptotically free region of moduli space.
Near $u = \infty$, the spectrum of the theory includes very heavy
BPS states satisfying $M_\text{m} \gg M_W \gg \La$.
Suppose the temperature $T \ll M_W$.
In this regime, it is valid to use an effective description of the low 
energy physics in terms of the scalar field $a(x)$ discussed in
Sec.~\ref{sec:review}. The appropriate effective theory
contains only the $U(1)$ vector multiplet $\A$ and is
described by 
the Lagrangian $\L_\text{eff}$ given in Eq.~\eqref{Leffs}.

Although it
is possible to construct an effective theory incorporating the additional 
massive charged vector multiplets (see, for instance, Ref.~\cite{SW1}), 
the $W$ bosons have mass $M_W = \sqrt{2}|a|$ which, by assumption,
is large compared to $T$. 
The Boltzmann weight of these $W$ bosons exponentially suppress their 
contribution to the free energy density relative to the contributions of 
the massless particles. Since the $W$ bosons and their superpartners form a 
dilute gas, it is straightforward to obtain%
\footnote{
    Each charged degree of freedom in the plasma generates a contribution
    to the free energy density equal to
    $f_\pm = \pm \half T\sum_{n\in\Z} \int\frac{d^3p}{(2\pi)^3} \>
    \ln\bigl[(\w_n^\pm)^2 + \vec{p}^{\,2} + M_W^2\bigr]$,
    where $\w_n^+ \equiv 2n\pi T$ for bosons, and $\w_n^- \equiv (2n{+}1)\pi T$ 
    for fermions.
    %%% and $n \in \Z$.
    The mass $M_W$ is the same for
    all particles belonging to a given multiplet. 
    Note that $M_W$ cuts off the infrared divergence in the spatial momentum
    integral of the bosonic zero-frequency contribution.
    The temperature-dependent part of the free energy contribution $f_\pm$
    is finite, and is a standard integral in statistical mechanics,
    $f_\pm = (T\text{-indep.}) \pm \frac{T^4}{2\pi^2} \int_0^\infty dx\, x^2 
    \ln\bigl(1 \mp e^{-\sqrt{x^2 + M_W^2/T^2}}\bigr)$.
    In a charged vector multiplet, the on-shell bosonic fields 
    (gauge field and complex scalar) account for four real degrees of freedom.
    By supersymmetry the on-shell fermionic fields (two Weyl fermions) must 
    also account for four real degrees of freedom. There are two such charged 
    multiplets for $SU(2)$ broken to $U(1)$. Thus, $(F/V)_\text{charged} = 
    8\,(f_+ + f_-)$. 
    Evaluating the leading contribution from the saddle-point at $x=0$
    immediately gives the result \eqref{FEcharged}.
}
\begin{equation}
\label{FEcharged}
    (F/V)_\text{charged} \approx
    -16 T^4\Bigl(\frac{M_W}{2\pi T}\Bigr)^{3/2}e^{-M_W/T}.
\end{equation}
Since the pressure is just minus the free energy density
(when all chemical potentials vanish),
the negative sign in this result shows that the dilute gas of heavy
particles exerts a positive pressure, as it must.
Eq.~\eqref{FEcharged} also indicates that the pressure decreases as $M_W/T$ grows
large. 
This may be achieved, at fixed temperature,
by moving toward large $|u|$ in the free energy surface. Also, note that the 
contributions of monopoles and dyons to the free energy will have even greater
exponential suppression since these excitations, for large $|u|$,
are much heavier than $W$ bosons.
This follows from the mass ratio of electrically and magnetically charged 
BPS-saturated states near infinity on moduli space,
$M_\text{m}/M_W = |a_D|/|a| \sim |\ln(u/\La^2)|$.
Thus, as noted in Ref.~\cite{Wirstam},
one may ignore massive charged fields altogether and focus on the 
interactions of the massless neutral fields.

To compute the effective potential $V_\text{eff}$ for the scalar field 
$a$, one may first integrate out thermal fluctuations of the vector
multiplet fields in perturbation theory.
As usual, one considers the functional integral representation 
for the partition function, with a constant source coupled to the scalar field,
and performs a saddle point expansion.
Let $a_0$ represent the translationally invariant expectation value for the
scalar field in the presence of the source. By construction, $a_0$ is a
solution to the Euler-Lagrange equations for a fixed value of the source. 
Let
\begin{equation}
\label{expansion}
    a(x) = a_0 + \atilde(x).
\end{equation}
Substituting this decomposition into the Lagrange density $\L_\text{eff}^{n=2}$
[given by expression \eqref{Leff2c}]
and using the prepotential \eqref{prepot} leads to
\begin{equation}
\label{Leff2cexpanded}
    \L_\text{eff}^{n=2} = 
    \L^{(0)}_\text{free} + \L^{(1)}_\text{int} + \L^{(2)}_\text{int} + \dotsb.
\end{equation}
The coupling constant $g_0^2 = 4\pi^2/(\ln|a_0/\La|^2+3)$
is small in the asymptotic regime under consideration.
The perturbative expansion for $V_\text{eff}$ will be a series controlled
by $g_0$.
The free Lagrange density is
\begin{equation}
    \L^{(0)}_\text{free} = 
    |\p_\mu\atilde|^2 + \fourth F_{\mu\nu}^2 + 
    \bigl(\tfrac{i}{2}\psi\s^\mu_\text{E}\p_\mu\psibar
    + \tfrac{i}{2}\la\s^\mu_\text{E}\p_\mu\labar
    + \text{H.c.}\bigr) - |F|^2 - \half D^2.
\label{eq:L0free}
\end{equation}
The interactions cubic and quartic in fluctuations are given by
\begin{subequations}
\begin{align}
    \frac{4\pi^2}{g_0^3}\,\L^{(1)}_\text{int} &= 
    \Bigl(\frac{\atilde}{a_0}+\frac{\atilde^*}{\abar_0}\Bigr)
    \L^{(0)}_\text{free} + \frac{i}{4}
    \Bigl(\frac{\atilde}{a_0}-\frac{\atilde^*}{\abar_0}\Bigr)F_{\mu\nu}
    \widetilde{F}_{\mu\nu} 
    + \Bigl(\frac{1}{a_0}\,\O_\text{fermi} + \text{H.c.}\Bigr), \\
    \frac{4\pi^2}{g_0^{4}}\, \L^{(2)}_\text{int} &= 
    - \frac{1}{2}\Bigl(\frac{\atilde^2}{a_0^2}+\frac{\atilde^{*2}}{\abar_0^2}\Bigr)
    \L^{(0)}_\text{free} - \frac{i}{8}
    \Bigl(\frac{\atilde^2}{a_0^2}-\frac{\atilde^{*2}}{\abar_0^2}\Bigr)
    F_{\mu\nu}\widetilde{F}_{\mu\nu}
    + \Bigl(-\frac{\atilde}{a_0^2}\,\O_\text{fermi} + \frac{1}{4a_0^2}\,\la^2\psi^2
    + \text{H.c.}\Bigr),
\end{align}
\label{eq:Lint}%
\end{subequations}
where
\begin{equation}
    \O_\text{fermi} \equiv
    -\tfrac{i}{2}\bigl(\psi\s_\text{E}^\mu\psibar + 
    \la\s_\text{E}^\mu\labar\bigr)\p_\mu\atilde
    + \half\psi^2 F^* + \half \la^2 F - \tfrac{i}{\sqrt{2}}
    \la\psi D - \tfrac{1}{\sqrt{2}}\la\s_\text{E}^{\mu\nu}\psi F_{\mu\nu}
\end{equation}
is a dimension five operator composed of fermion bilinears.
To obtain Eqs.~\eqref{eq:L0free} and \eqref{eq:Lint},
all fluctuating fields have been rescaled by a common factor of $g_0$.
In general, the
contribution $\L_\text{int}^{(p)}$ to the interaction Lagrange
density involves $2+p$ factors of fluctuating fields.
Each such term has
an overall factor of $g_0^{p+2}/|a_0|^p$
multiplying operators of dimension $4+p$.%
\footnote
    {
    The reason $\L_\text{int}^{(p)}$, for $p>0$,
    contains an overall factor of $g_0^{p+2}$ (instead of just $g_0^p$)
    is due to the fact that in the original expression \eqref{Leff2c} for 
    $\L_\text{eff}^{n=2}$, the K\"ahler connection $\Gamma(a,\bar a)$ and
    $R(a,\bar a)$ contain an inverse power of the K\"ahler metric
    $\gamma(a,\bar a)$ which (because $\gamma$ coincides with $g_{\rm eff}^{-2}$)
    cancels the overall factor of $g_{\rm eff}^{-2}$.
    }

A schematic expression for the effective potential is
\begin{equation}
\label{Veffseries}
\begin{split}
    V_\text{eff} &=
    -\frac{\pi^2}{12}\,T^4
    + \Bigl\langle \L_\text{int}^{(1)} \Bigl\rangle_0^\text{1PI}
    + \Bigl\langle \L_\text{int}^{(2)}
    + \bigl(\L_\text{int}^{(1)}\bigr)^2 \Bigr\rangle_0^\text{1PI}
    + \Bigl\langle \L_\text{int}^{(3)}
    + \L_\text{int}^{(1)}\,\L_\text{int}^{(2)}
    + \bigl(\L_\text{int}^{(1)}\bigr)^3 \Bigr\rangle_0^\text{1PI} \\
    &\quad 
    + \Bigl\langle \L_\text{int}^{(4)} 
    + \L_\text{int}^{(1)}\,\L_\text{int}^{(3)}
    + \bigl(\L_\text{int}^{(2)}\bigr)^2 
    + \bigl(\L_\text{int}^{(1)}\bigr)^2\L_\text{int}^{(2)}
    + \bigl(\L_\text{int}^{(1)}\bigr)^4 \Bigr\rangle_0^\text{1PI} + \dotsb.
\end{split}
\end{equation}
The first term in Eq.~\eqref{Veffseries} is the blackbody contribution from a
massless $\Ntwo$ vector multiplet.%
\footnote{
    Note that there are no background-dependent terms in
    $\L^{(0)}_{\rm free}$.
    Consequently, the one-loop contribution to $V_\text{eff}$
    (involving the logarithm of a functional determinant)
    simply gives the blackbody result.
}
In the remaining terms, we have omitted
spacetime integrals and combinatorial coefficients.
A generic term of the form 
$\vev{\prod_{i=1}^m\L_\text{int}^{(p_i)}}_0^\text{1PI}$ represents
the expectation value of $m$ spacetime integrals of the interaction Lagrange
densities in the unit-normalized Gaussian measure. If this term is expressed
diagrammatically, then each diagram will have $m$ vertices (representing
insertions of the $\L_\text{int}^{(p_i)}$), joined by propagators arising
from $\L_\text{free}^{(0)}$. Only the one-particle irreducible (1PI)
portion of these correlators contributes to the effective potential.

The effective potential admits a double series expansion in the dimensionless
coupling $g_0$ and the ratio of scales $T/|a_0|$.
To this end, the expansion in Eq.~\eqref{Veffseries}
has been organized by operator dimension, with
the dimension 5, 6, 7, and 8 terms shown explicitly.
The Gaussian measure is invariant under independent $U(1)$ phase rotations
for $\atilde$, $\psi$, $\la$, and 
$F$, and $\Z_2$ parity transformations for $A_\mu$ and $D$. 
These symmetries immediately imply that the dimension 5 and 7 terms in
Eq.~\eqref{Veffseries} (which come with odd powers of $g_0$), and
the term $\bigl\langle\L_\text{int}^{(2)}\bigr\rangle_0^\text{1PI}$,
vanish identically. 

From $\bigl\langle\bigl(\L_\text{int}^{(1)}\bigr)^2\bigr\rangle_0^\text{1PI}$ 
there are eight basic diagrams that must be considered. 
These ``basketball diagrams'' are shown in Figure \ref{fig:holoinfty}.%
\footnote{
    Although auxiliary field propagators are momentum-independent, the diagrams
    containing $F$ and $D$ propagators have two independent loop momenta. 
    This is clear if the diagrams are constructed using position-space Feynman
    rules. Diagrams \ref{fig:holoinfty}c and \ref{fig:holoinfty}d may be thought
    of as originating from four-fermion interactions in the on-shell 
    formalism. The auxiliary field is a constraint
    that causes these graphs to ``pinch,'' yielding a graph with a
    figure-eight topology.
    Such an on-shell diagram has two fermion loops which
    contribute two overall factors of $-1$; this is matched in the off-shell 
    diagram by one fermion loop and one central auxiliary field propagator
    that each contribute a factor of $-1$.
}
All of these diagrams vanish because they reduce to the sum-integrals%
\footnote{
    Sum-integrals are defined as 
    $\inlinesumint{\ell}{\pm} = T \sum_{\ell_0}\int\!\!\frac{d^3\ell}{(2\pi)^3},$
    where the sum is over even ($+$) or odd ($-$) integer multiples of $\pi T$.
}
\begin{subequations}
\label{sumints}%
\begin{align}
\label{sumintsA}
    \sumint{p}{\pm}\sumint{q}{\pm} & \> \frac{p\cdot q}{p^2 \, q^2} = 0\,,
\\
    \noalign{\noindent\text{or}}
    \sumint{p}{\pm} & 1 = 0\,.
\label{sumintsB}
\end{align}
\end{subequations}
Expression \eqref{sumintsA} 
vanishes by Euclidean time reflection
and spatial parity invariance. To justify this one may choose to regulate the theory
by dimensional continuation, which preserves spacetime symmetries.
[Or one may ignore the issue of regulation, since that is 
part of defining the theory at zero temperature,
and focus only on the temperature-dependent part of the sum-integral.
Each discrete frequency sum may be recast as a 
pair of contour integrals just above the real axis: one temperature-independent
and the other temperature-dependent. 
The latter contains the appropriate statistical 
distribution function which dies off exponentially fast in the upper half 
complex plane. The temperature-dependent piece is finite and 
vanishes by the symmetry arguments.]
Expression \eqref{sumintsB}
has a scale-free spatial momentum integral that vanishes in 
dimensional continuation. [Alternatively, in the contour integral method,
the temperature-dependent integrand is analytic in the upper half plane, so
closing the contour there produces zero.] Thus, there is no 
$O(g_0^6 \, T^2/|a_0|^2)$ contribution to $V_\text{eff}/T^4$. 
This is consistent with the analysis of Ref.~\cite{Wirstam}.

\FIGURE[t]{
    \label{fig:holoinfty}
    \centerline{\includegraphics[width=4.0in]{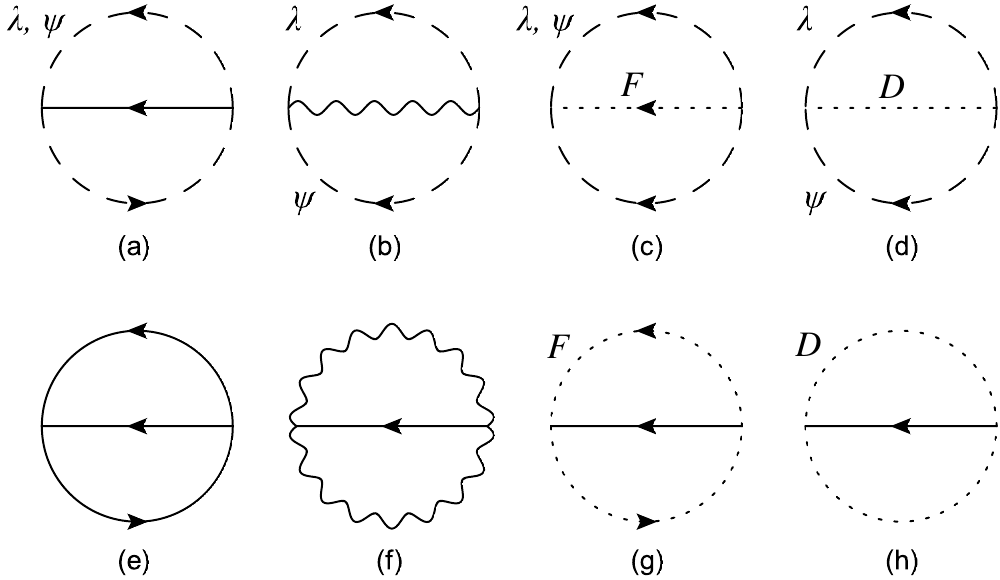}}
    \vspace*{-20pt}
    \caption{Two-loop 1PI diagrams that could contribute to the effective 
    potential at $O(g_0^6 T^6/|a_0|^2)$. Solid lines represent complex scalar 
    fields, dashed lines represent either type of Weyl fermion, wavy lines 
    represent Abelian gauge fields, and dotted lines represent auxiliary 
    fields.}
}

Continuing on to the subsequent terms in the expansion \eqref{Veffseries}
for the effective potential, one finds that contributions
to $V_\text{eff}/T^4$ from dimension 8 operators debut at $O(g_0^8\, T^4/|a_0|^4)$.
The local dimension 8 piece
$\langle\L_\text{int}^{(4)}\rangle_0$
displayed in the Eq.~\eqref{Veffseries}
has an overall factor of $g_0^6$
but (just as for $\langle\L_\text{int}^{(2)}\rangle_0$
this piece vanishes due to phase rotation symmetries.

It was pointed out in Ref.~\cite{Wirstam} that the first non-vanishing
correction to the 
effective potential does not come from the $n\,{=}\,2$ terms in the low 
energy effective action, but rather from the higher derivative $n\,{=}\,4$ terms.
In other words, $\L_\text{eff}^{n=4}$ and the non-holomorphic function $\K$ is more
important than $\L_\text{eff}^{n=2}$ and the holomorphic prepotential $\F$, when
it comes to understanding the leading dependence of the effective potential on 
the expectation value of the scalar field. To see this one must use the known 
form of $\K$ \cite{deWit,Lind},
\begin{equation}
\label{K}
    \K(A,\Abar) \approx 
    \frac{c}{64} %c
    \ln\Bigl(\frac{A^2}{\La^2}\Bigr)
    \ln\Bigl(\frac{\Abar^2}{\La^2}\Bigr),
\end{equation}
with $c$ a constant.
This is the leading approximation for the non-holomorphic function $\K$ when
$a \equiv A|_{\th = \thbar = 0} \gg \La$.%
\footnote
    {%
    This term is responsible for producing
    four derivative terms in the effective action, such as $F^4$.
    (It is simpler to discuss perturbative
    corrections to $\int d^4\th \, d^4\thbar\>\K$, rather than $\K$ itself,
    since the former is a Lagrange density and thus a direct output
    of renormalizing the short distance theory.)
    It has been shown, at least in the case of $\Ntwo$ QED,
    that there is a nonvanishing quantum correction at two-loop order
    to the supersymmetric completion of the $F^4$ term \cite{KuzMc1}.
    In our case, the $F^4$ term is given
    by $-4C\int d^2\th \, d^2\thbar\> \K_{AA\Abar\Abar}\,W^2\Wbar^2$, and the
    corresponding statement about quantum corrections to this operator may
    be expressed as $C = 1 + O(g_0^2)$.  The loop expansion parameter
    in the short distance theory is the running coupling $g_0^2$,
    evaluated at the scale $a$, or in other words, the inverse logarithm
    $(\ln |a/\La|^2)^{-1}$.
    Far out in moduli space, this coupling is small.
    Such two-loop corrections lead to $O(g_0^6 T^8/|a_0|^4)$ contributions
    to the effective potential which will turn out to be subleading.
    }
A couple points regarding the formula \eqref{K} for $\K$ are worth 
mentioning. First, the concise form \eqref{K} is due to the fact that the 
low energy gauge group is Abelian.%
\footnote{
    The leading form of $\K$ for non-Abelian gauge group $SU(2)$ was originally 
    determined in Ref.~\cite{deWit}. On the Coulomb branch of $SU(2)$ $\Ntwo$ theory,
    the leading form of $\K$ simplifies considerably, as noted in Ref.~\cite{Lind}.
}
Second, both $U(1)_R$ and scale transformations (of the form $\La \to b\La$) 
change Eq.~\eqref{K} additively by terms that are (anti)holomorphic in the 
chiral superfield. Such terms vanish under the full superspace integration,
and therefore have no effect on the dynamics.

For simplicity, we compute the contribution to the free energy density from 
the purely scalar terms in $\L_\text{eff}^{n=4}$. Such terms originate from
the piece
\begin{equation}
\label{nonholoscalar}
    \L_\text{eff}^{n=4} \supset -\int d^2\th \, d^2\thbar\>
    \K_{A\Abar}(A,\Abar)\left[
    (D^\a D_\a A)(\Dbar_\adot \Dbar^\adot \Abar)
    + 2(\Dbar_\adot D^\a A)(D_\a \Dbar^\adot \Abar)
    \right].
\end{equation}
We write expression \eqref{nonholoscalar} in components, perform the expansion
\eqref{expansion}, and rescale fluctuating fields by $g_0$.%
\footnote{
    When chiral superfields occur in denominators, we factor out the 
    lowest (scalar) component and expand the remaining function as a power 
    series in $\th$ and $\thbar$.
}
The resulting purely scalar terms that are both invariant
under phase rotations of the fluctuation $\tilde a$,
and suppressed by no more than $|a_0|^{-4}$, are given by%
\footnote
    {
    Contributions from multi-point correlators of terms which are
    not individually invariant under phase rotations of $\tilde a$
    first contribute to the 
    effective potential at $O(g_0^6 T^8/|a_0|^4)$.
    This is smaller by two powers of $g_0$ than
    the contributions which will result from the unperturbed expectation
    of the $U(1)$-invariant terms \eqref{purescalar}.
    }
\begin{equation}
\label{purescalar}
\begin{split}
    \L_\text{eff}^{n=4}
    \supset
    -c \, \biggl[
    \frac{g_0^2}{|a_0|^2}\bigl(|\p_\mu\p_\nu\atilde|^2 
    + |\p^2\atilde|^2\bigr)
    +\frac{g_0^4}{|a_0|^4}
    & \Bigl\{
    (|\p_\mu\atilde|^2)^2 
    + |\atilde|^2 |\p_\mu\p_\nu\atilde|^2 + |\atilde|^2|\p^2\atilde|^2
\\
    &+ 
    \bigl[
	\atilde\,(\p_\mu\p_\nu\atilde)(\p_\mu\atilde^*)(\p_\nu\atilde^*) + 
	\text{H.c.}
    \bigr] \Bigr\} \biggr] \,,
\end{split}
\end{equation}
up to total spacetime derivatives. It will turn out 
that only the first two terms inside the curly braces generate a non-zero 
contribution to $V_\text{eff}/T^4$ of $O(g_0^4 T^4/|a_0|^4)$.

\FIGURE[t]{
    \label{fig:feynrule}
    \centerline{\includegraphics{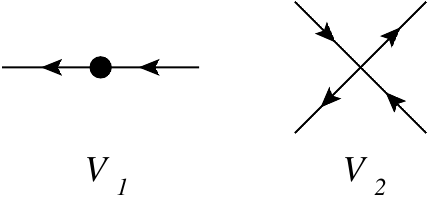}}
    \vspace*{-20pt}
    \caption{Vertices representing leading four-derivative scalar self-interactions.}
}

The interaction terms \eqref{purescalar} generate
quadratic and quartic interaction vertices,
illustrated in Fig.~\ref{fig:feynrule},
with momentum-dependent vertex factors.
Explicitly, the resulting (Euclidean space)
vertex factors are
\begin{subequations}
\begin{align}
    V_1
    &= \frac{2c g_0^2}{|a_0|^2} \, (p^2)^2, \\
    V_2
    &= \frac{c g_0^4}{|a_0|^4}
    \Bigl[2(p_1\cdot p_2)(p_3\cdot p_4) + 2(p_1\cdot p_4)(p_2\cdot p_3) 
    + p_1^2\, p_2^2 + p_2^2\, p_3^2 + p_3^2\, p_4^2 + p_4^2\, p_1^2
    \nonumber \\
    &\qquad
    + (p_1\cdot p_2)^2 + (p_2\cdot p_3)^2 + (p_3\cdot p_4)^2 + (p_4\cdot p_1)^2
    \label{quarticrule} \\
    &\qquad
    + 2(p_1\cdot p_2)(p_1\cdot p_4) + 2(p_2\cdot p_3)(p_3\cdot p_4) 
    + 2(p_1\cdot p_2)(p_2\cdot p_3) + 2(p_1\cdot p_4)(p_3\cdot p_4)
    \Bigr]. \nonumber
\end{align}
\end{subequations}
The momenta are taken along the arrows (which distinguish $\atilde^*$ from 
$\atilde$).
The overall signs of both vertex factors are positive 
because there is a minus sign from the definition of the Euclidean path 
integral weight, a minus sign from the Lagrange density itself, and two factors
each of $i$ and $-i$ from Fourier transforms of derivatives.

\FIGURE[t]{
    \label{fig:nonholoinfty}
    \centerline{\includegraphics[width=2.5in]{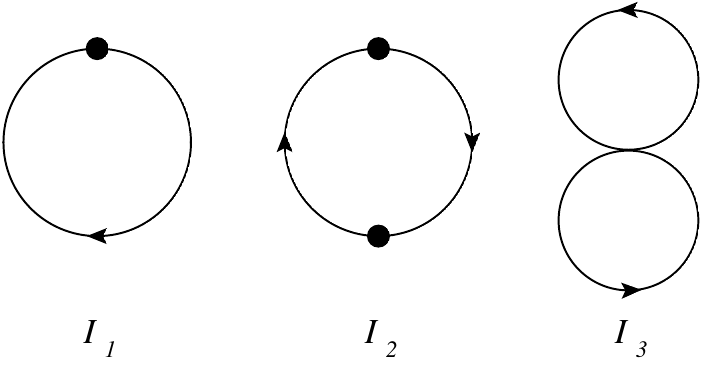}}
    \vspace*{-20pt}
    \caption{Leading scalar contributions to free energy density up to 
    $O(g_0^4 T^8/|a_0|^4)$.}
}

These quadratic and quartic vertices generate the
bubble diagrams shown in Figure \ref{fig:nonholoinfty}.
Diagrams $I_1$ and $I_2$ vanish,
since their spatial momentum integrals are scale-free.
The last diagram, $I_3$, reduces to the square of a nontrivial one-loop 
sum-integral,%
\footnote{
The last eight terms in the Feynman rule for $V_2$ in
    Eq.~\eqref{quarticrule} contribute terms to the expression for $I_3$
    that integrate to zero. These eight terms 
    originate from the operators $|\atilde|^2|\p^2\atilde|^2$ and 
    $\atilde(\p_\mu\p_\nu\atilde)(\p_\mu\atilde^*)(\p_\nu\atilde^*) 
    + \text{H.c.}$ in expression \eqref{purescalar}.
    Since these operators have no bearing on the calculation,
    they are omitted in Eq.~(3.6) of Ref.~\cite{Wirstam}.
}
\begin{equation}
\begin{split}
\label{I3}
    I_3
    &= \b V \, \frac{1}{2} \sumint{p}{+}\sumint{q}{+} 
    \frac{c\, g_0^4}{|a_0|^4}
    \biggl[\frac{4(p\cdot q)^2}{p^2q^2}\biggr]
    = \b V \, \frac{8 c\, g_0^4}{3|a_0|^4}
    \biggl(\sumint{p}{+}\frac{\vec{p}^{\,2}}{p^2}\biggr)^2,
\end{split}
\end{equation}
where we have used rotational symmetry to replace 
$\frac{(p\cdot q)^2}{p^2 q^2}$ by 
$(1+\frac{1}{d-1})\frac{\vec{p}^{\,2}\vec{q}^{\,2}}{p^2q^2}$
inside the integrals and $d$ is the spacetime dimension
(continued infinitesimally away from $4$). 
In Eq.~\eqref{I3} the sum-integral in parentheses
$J_+ \equiv \inlinesumint{p}{+}\frac{\vec{p}^{\,2}}{p^2}$
evaluates to $\pi^2 T^4/30$. The scalar
self-interactions contribute $-(\b V)^{-1}I_3$ to the effective potential,
so
\begin{equation}
    V_\text{eff}(a_0)\bigr\vert_\text{scalar-scalar} = 
    -c\,\frac{2\pi^4}{675} \frac{g_0^4 T^4}{|a_0|^4} \,T^4.
\end{equation}
Obviously,
the sign of the coefficient $c$ is of the utmost importance for 
interpreting the effective potential --- the sign determines whether local 
equilibrium is directed toward smaller or larger values of $|a_0|$. 

An attempt has been made to derive the non-holomorphic function $\K$ for 
$SU(2)$ $\Ntwo$ gauge theory and fix the value of the overall constant \cite{Ketov}.
The method involves technical superfield calculations and is difficult to check
for errors.%
\footnote{
    In the case of $SU(2)$ $\Nfour$ gauge theory, three independent superfield
    calculations of the non-holomorphic function $\K$ are known, and they
    completely agree \cite{GonzRey,Buchbind,Tseytlin}. The calculations for the
    superconformal theories discussed in these works may be readily extended to 
    $SU(2)$ $\Ntwo$ gauge theory. For example, Eq.~(4.9) in Ref.~\cite{Tseytlin}
    represents the one-loop effective action for $\Ntwo$ gauge theory with
    four fundamental hypermultiplets and gauge group $SU(2)$ Higgsed to $U(1)$.
    Of direct relevance to this work is the sum of the second and third terms on 
    the right hand side of Eq.~(4.9)---this coincides with the non-holomorphic 
    contribution to the effective action for $\Ntwo$ theory without matter.
    The $F^4$ term may be extracted by Taylor expanding the functions 
    $\zeta(t\Psi,t\bar{\Psi})$ and $\w(t\Psi,t\bar{\Psi})$ to zeroth order around
    $\Psi = \bar{\Psi} = 0$. This amounts to focusing on just the leading term in a
    derivative expansion. Since $\zeta(0,0) = 1/12$ and $\w(0,0) = 0$, it follows
    that $c$ is positive. A positive value for $c$ agrees with the conclusion in
    this work. We thank the JHEP referee for explaining this to us.
}
Our goal, in the next
couple of pages, is to find an independent determination of the sign of $c$,
since that is the crucial information needed to understand low temperature 
thermodynamics in this theory.

A simple and physical method that fixes the sign of the coefficient $c$ 
is provided by studying the forward amplitude for scalar scattering 
\cite{Adams}. There are spinless, one-particle states in the spectrum of the
theory that arise from quantum fluctuations of the field $\atilde$. 
Let us denote the particle excitation by $\varphi$ and its 
antiparticle by $\bar{\varphi}$. Consider the scattering process 
$\varphi\bar{\varphi} \to \varphi\bar{\varphi}$.
At center-of-momentum energies
far below $M_W$, one may use the low energy effective action to reliably 
compute the scattering amplitude in a momentum expansion. The tree-level 
diagram for this process is derived entirely from the vertex $V_2$ shown 
in Figure \ref{fig:feynrule}, remembering that the Minkowski space vertex gets 
an extra factor of $i$ relative to its Euclidean counterpart.%
\footnote{
    For scattering we take all momenta as incoming (rather than following 
    the arrows) and label them as 
    $p_i$, $i = 1,\dotsc,4$ starting at the upper left corner and 
    continuing counterclockwise.
    This amounts to flipping the signs of $p_2$ 
    and $p_4$ in Eq.~\eqref{quarticrule}, which clearly does nothing to the
    whole expression.
    We use a Minkowski space with $-+++$ signature,
    and define the usual Mandelstam variables 
    $s = -(p_1+p_2)^2$ and $t = -(p_2+p_3)^2$.
    The mass-shell condition is $p_i^2 = 0$.
}
It is worth noting that, from the point of view of the expanded low energy 
effective Lagrange density,
the operators appearing in expression \eqref{purescalar} 
comprise only a subset of the irrelevant interactions.
Moreover, they are not even the ones with lowest dimension.
Nevertheless, the quartic operators in
expression \eqref{purescalar} generate the leading
contribution to the
$\varphi\bar{\varphi} \to \varphi\bar{\varphi}$ scattering process.
Contributions from other terms in the effective theory are suppressed
by additional factors of the dimensionless coupling $g_0$.
The resulting Lorentz-covariant scattering amplitude is%
\footnote{
    Recall that the LSZ reduction formula relates the Fourier transform of 
    $i$ times time-ordered correlation functions to the 
    scattering amplitude $\M$. Thus, $-i\M$ is the object to which 
    diagrammatic rules apply.
}
\begin{equation}
    -i\,\M(s,t) = i\,\frac{c g_0^4}{|a_0|^4}\,(s+t)^2 + O(g_0^6).
\end{equation}
In the forward scattering limit,
\begin{equation}
\label{forward}
    \A(s) = \lim_{t \to 0^-}\M(s,t) 
    = -c \, g_0^4 \frac{s^2}{|a_0|^4} + O(g_0^6).
\end{equation}

Now consider the contour integral
\begin{equation}
    I = \oint_\g\frac{ds}{2\pi i}\> \frac{\A(s)}{s^3}\,.
\end{equation}
The analytic structure of the exact forward scattering amplitude in the 
complex $s$-plane is shown in Figure \ref{fig:contour}. $\A(s)$
must have a branch cut along the positive, real $s$-axis with a branch point
that corresponds to the threshold for pair production.
Since $\Ntwo$ gauge 
theory has excitations with arbitrarily low momentum, there is no mass gap and
the branch point sits at the origin.
Following Ref.~\cite{Adams}, one may
modify the theory in the deep IR by giving a small regulator mass 
$m_\text{gap}$ to the $\atilde$ fields. The cut then extends only down
to $(2m_\text{gap})^2$.
There is no cut along the negative real axis,
since the amplitude $\M$ is only symmetric under interchange
of $s$ and $t$, not under interchange of $s$ and $u$.
By construction,
the integrand $\A(s)/s^3$ also has a pole at the origin.

\FIGURE[t]{
    \label{fig:contour}
    \centerline{\includegraphics{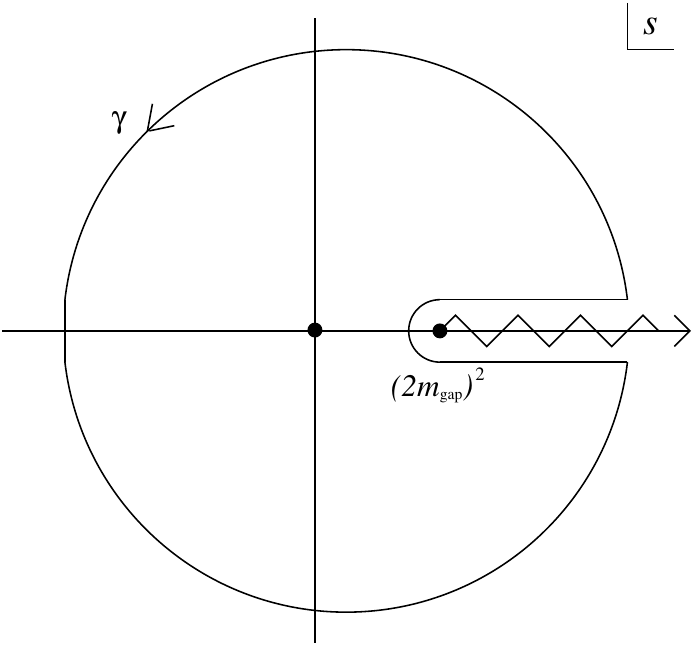}}
    \vspace*{-20pt}
    \caption{Analytic structure of the forward amplitude $\A(s)$ as a function 
    of (complexified) center-of-momentum energy squared.
    $\Ntwo$ gauge theory is recovered by sending $m_\text{gap} \to 0$.}
}

The integral $I$ may be evaluated by deforming the contour $\g$ usefully 
in one of two ways: in a 
tight circle around the origin yielding the residue at the pole at the origin,
or around
infinity. In the latter scenario the integral along the large circular portion 
of the contour vanishes since $SU(2)$ $\Ntwo$ gauge theory is UV-complete and 
thus the forward amplitude (when the theory has a mass gap) 
grows no faster than $s^2$ at high energies. This leaves
only the contour wrapping the cut which measures the integrated discontinuity
of $\A(s)$ across the cut. Since $\A(s)$ is real along part of the real 
$s$-axis, the Schwarz reflection principle relates the discontinuity to the 
imaginary part of $\A(s)$ just above the cut. Consequently,
\begin{equation}
\label{contoureq}
    \tfrac{1}{2} \, \A''(0) = 
    \frac{1}{\pi}\int_{4m_\text{gap}^2}^\infty ds \>
    \frac{\Im[\A(s+i\e)]}{s^3}\, .
\end{equation}
On the left-hand side of Eq.~\eqref{contoureq}, one may approximate the forward
amplitude at weak coupling with the tree-level formula. Since the introduction 
of a gap modifies the mass-shell condition, the tree-level amplitude is given 
by Eq.~\eqref{forward} with additional terms of 
$O(m_\text{gap}^4)$ or $O( m_\text{gap}^2\, s)$.
However, these additional terms have no
effect on the final result, since only the unmodified $s^2$ behavior is
extracted by the residue theorem.
Because the tree-level amplitude Eq.~\eqref{forward} is analytic at
the origin,
%% (low energy cuts are absent at leading order in the coupling),
we find $\half \A''(s=0) = -c\, g_0^4/|a_0|^4$.
The right-hand side of Eq.~\eqref{contoureq} involves an integral
of a negative-definite quantity,
as unitarity of the $S$-matrix requires that $\Im[\A(s+i\e)] < 0$.%
\footnote{
    One may phrase the fact that the imaginary part of the forward amplitude 
    must be negative %%(in our choice of metric)
    in terms of the optical theorem,
    as done in Ref.~\cite{Adams}. Ultimately, the negative sign reflects
    $S$-matrix unitarity, since
    $S = 1 - i\M$ and 
    $S S^\dag = 1$ imply that $\Im(\M) = -\half|\M|^2 < 0$. 
    See, for example, problem 17 in Ch.~3 of Ref.~\cite{Brown}.
}
Hence, $c$ must be positive.%
\footnote{
    As discussed in Ref.~\cite{Adams}, this constraint on the sign of $c$ is a 
    special case of a more general scenario. The forward amplitude $\A(s)$, 
    away from the real axis and probing $m_\text{gap}^2 \ll |s| \ll M_W^2$, 
    has a Taylor expansion around any point $s_0$ in this region that begins 
    as $(s-s_0)^2$ with a coefficient that is negative, up to corrections 
    which scale as $O(|s_0|^2/M_W^2, m_\text{gap}^2/M_W^2)$. In our case, the
    low energy theory is weakly-coupled and this permits an analytic expansion
    for $\A(s)$ at the origin.
}

We conclude that the scalar self-interaction provides a negative contribution
to the effective potential at $O(g_0^4\, T^8/|a_0|^4)$. There are, of course,
additional interactions from $\L_\text{eff}^{n=4}$ that should also be
considered.
The local operators whose thermal expectation values lead to
$O(g_0^4 T^8/|a_0|^4)$ corrections to the effective potential are all
dimension eight and involve four fields.
They give rise to the set of two-loop diagrams shown in Figure 
\ref{fig:twoloop}.
The scalar-scalar contribution $I_{\text{ss}}$ in
Fig.~\ref{fig:twoloop} is the just-discussed $I_3$.
The other diagrams all have the same figure-eight topology and involve
some pairing of scalars, fermions, and vectors running in the two loops.
They were calculated in Ref.~\cite{Wirstam},
and were found to all have the same relative sign.
However, the sign of these contributions asserted in Ref.~\cite{Wirstam}
is opposite to our conclusion for $I_{\text{ss}}$,
stemming from the fact that the single overall coefficient
$c$ is negative in Ref.~\cite{Wirstam}.%
\footnote{
\label{cneg}%
    The negativity of $c$ in Ref.~\cite{Wirstam} may be traced back to
    one of the references cited in that paper. The value of $c$ may be 
    obtained from Eq.~(3.11) of Ref.~\cite{Ketov}; expressing their
    equation in the form of our Eq.~\eqref{Leffss} implies that
    $c = -1/(8\pi^2)$, which has the wrong sign.
}
We now review the calculations of these diagrams and adjust the conclusions 
in lieu of the fact that $c$ is actually positive.

\FIGURE[t]{
    \label{fig:twoloop}
    \centerline{\includegraphics{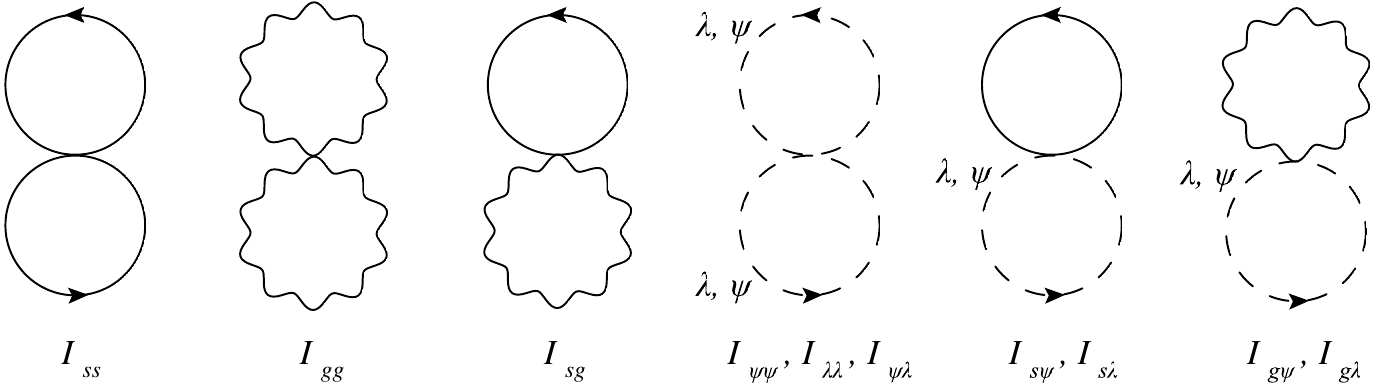}}
    \vspace*{-20pt}
    \caption{Two-loop diagrams that contribute to the effective potential at 
    $O(g_0^4 T^8/|a_0|^4)$. Solid lines represent complex scalar fields, dashed
    lines represent either type of Weyl fermion, and wavy lines represent 
    Abelian gauge fields.}
}

The $F^4$ terms of the low energy effective Lagrange density are readily 
obtained from the last term in expression \eqref{Leff4s} involving four 
copies of the spinor-valued field strength.
The purely gauge interactions that are 
suppressed by no more than $|a_0|^{-4}$ are
\begin{equation}
    \L_\text{eff}^{n=4} \supset 
    - \frac{c g_0^4}{16|a_0|^4}\Bigl[(F_{\mu\nu}F_{\mu\nu})^2 
    - (F_{\mu\nu}\widetilde{F}_{\mu\nu})^2\Bigr].
\end{equation}
The corresponding diagram in Figure \ref{fig:twoloop} is $I_\text{gg}$.
We find that (for an arbitrary Lorentz gauge-fixing parameter),
\begin{equation}
    I_\text{gg} = \frac{d}{2}\, \frac{c g_0^4}{|a_0|^4} \, \b V \, \Omega_{++}\,,
\end{equation}
where $\Omega_{++} \equiv \inlinesumint{p}{+}\inlinesumint{q}{+} 
\frac{(p\cdot q)^2}{p^2 q^2} = \frac{4}{3}J_+^2$ and $d = 4$. 
This is the same double sum-integral that we found in our evaluation
of the scalar self-interactions.
Note that $I_\text{gg} = I_\text{ss}$.

The four-fermion interaction with all $\psi$'s is easily derived from the same 
superfield integral used to obtain the purely scalar interactions, namely 
expression \eqref{nonholoscalar}. The purely $\psi$ interactions
that are suppressed by $|a_0|^{-4}$ are
\begin{equation}
    \L_\text{eff}^{n=4} \supset 
    -\frac{c g_0^4}{|a_0|^4}
    \bigl[(\psibar\sbar_\text{E}^\mu\p_\mu\psi)
    (\p_\nu\psibar\sbar_\text{E}^\nu\psi) 
    - (\psi\p_\mu\psi)(\psibar\p_\mu\psibar)\bigr].
\end{equation}
The four-fermion interaction with all $\la$'s must have exactly the same form
by $SU(2)_R$ symmetry. In Figure \ref{fig:twoloop}, the corresponding diagrams
are $I_{\psi\psi}$ and $I_{\la\la}$. We find
\begin{equation}
    I_{\psi\psi} = I_{\la\la} =
    2\,\frac{c g_0^4}{|a_0|^4}\,\b V \,\Omega_{--}\,,
\end{equation}
where $\Omega_{--} \equiv \inlinesumint{p}{-} \inlinesumint{q}{-} 
\frac{(p\cdot q)^2}{p^2 q^2} = \frac{4}{3}J_-^2$ and 
$J_- \equiv \inlinesumint{p}{-} \frac{\vec{p}^{\,2}}{p^2} = -7\pi^2 T^4/240$.

It is straightforward, but tedious, to find the interactions that mix scalars 
and vectors, or mix different types of Weyl fermion. To save ourselves some 
trouble we rely on the component field expression of $\L_\text{eff}^{n=4}$
given in Eq.~(3.6) of Ref.~\cite{Wirstam}. Based on our discussion in footnote
\ref{cneg}, we shall factor out an overall $-1/(8\pi^2)$
and replace it by the constant $c$. According to Ref.~\cite{Wirstam}, the 
scalar-gauge and $\psi$-$\la$ interactions that are
suppressed by $|a_0|^{-4}$ and contribute to finite temperature effects are 
given by%
\footnote{
    Our $\s$-matrix conventions differ from those of Ref.~\cite{Wirstam}.
    In particular, $\s_\text{E}^0(\text{ours})= -\s_\text{E}^0(\text{theirs})$,
    with similar spatial matrices. The two conventions are related by a spatial
    parity transformation which has the effect of conjugating $\s$-matrices by
    $\s^0$ (or $\sbar^0$ as appropriate). We have done a parity transformation
    in order to write the mixed $\psi$-$\la$ interactions in our convention.
}
\begin{equation}
    \L_\text{eff}^{n=4} \supset
    -\frac{2c g_0^4}{|a_0|^4}\bigl[
    (\p_\mu\atilde^*)(\p_\nu\atilde) F_{\mu\rho} F_{\nu\rho}
    -(\labar\p_\mu\psibar)(\psi\s_\text{E}^{\mu\nu}\p_\nu\la)
    -(\la\p_\mu\psi)(\psibar\sbar_\text{E}^{\mu\nu}\p_\nu\labar)\bigr].
\end{equation}
In Figure \ref{fig:twoloop}, the corresponding diagrams are $I_\text{sg}$ and 
$I_{\psi\la}$. We find
\begin{equation}
    I_\text{sg} = 2(d{-}2)\frac{c g_0^4}{|a_0|^4}\,\b V \,\Omega_{++},
    \qquad
    I_{\psi\la} = 4\,\frac{c g_0^4}{|a_0|^4}\,\b V \, \Omega_{--}.
\end{equation}
(Reassuringly, the gauge dependence cancels completely in $I_\text{sg}$.)

It remains to compute the diagrams for the scalar-fermion and gauge-fermion
interactions. In Figure \ref{fig:twoloop},
these contributions are $I_{\text{s}\psi}$, 
$I_{\text{s}\la}$, $I_{\text{g}\psi}$, and $I_{\text{g}\la}$. 
Since each vertex involves one type of Weyl 
fermion paired with its Hermitian conjugate, $SU(2)_R$ symmetry requires 
equivalent interactions for $\psi$ and $\la$. It follows that 
$I_{\text{s}\psi} = I_{\text{s}\la}$ and $I_{\text{g}\psi} = I_{\text{g}\la}$. 
The latter relation means that it is impossible for 
there to be any nontrivial gauge dependence in the diagrams involving gauge
fields since there are no other diagrams left to cancel it. Since the 
complex scalar and gauge fields belong to the same supersymmetry multiplet, 
they have equal numbers of propagating degrees of freedom. Hence, all four 
diagrams must be equal. To determine their common value consider computing the 
index $\tr\,((-1)^F e^{-\b H})$ at weak coupling. In perturbation theory, the 
$O(g_0^4)$ contribution to the index comes from the class of diagrams shown
in Figure \ref{fig:twoloop}, but with periodic temporal boundary conditions 
for all fields. This means that all frequency sums are taken over even integer
multiples of $\pi T$. Effectively, this changes all 
instances of $\Omega_{--}$ and $\Omega_{+-}$ to $\Omega_{++}$. The index, which
must be an integer, cannot change as the coupling $g_0$ is varied. Therefore, 
the $O(g_0^4)$ part of the index must be identically zero,
\begin{equation}
\begin{split}
    0 &= \bigl[
    I_\text{ss} + I_\text{gg} + I_{\psi\psi} + I_{\la\la} + I_\text{sg} 
    + I_{\psi\la} + I_{\text{s}\psi} + I_{\text{s}\la} + I_{\text{g}\psi} 
    + I_{\text{g}\la}\bigr]\Big|_\text{p.b.c.} \\
    &= 16\,\frac{c g_0^4}{|a_0|^4}\, \b V \, \Omega_{++} 
    + 4 I_{\text{s}\psi}\big|_\text{p.b.c.}.
\end{split}
\end{equation}
Consequently,
\begin{equation}
\label{Ispsi}
    I_{\text{s}\psi}\big|_\text{p.b.c.}
    = -4\,\frac{c g_0^4}{|a_0|^4}\, \b V\, \Omega_{++}\, .
\end{equation}
This is a contribution to the index, but what we really want is the 
contribution to the partition function $\tr\,(e^{-\b H})$. The functional 
representation of the trace requires antiperiodic temporal boundary conditions
for fermions. Since each diagram involves a single fermion loop, we only need 
to turn one of the frequency sums in Eq.~\eqref{Ispsi} into a sum over odd integer 
multiples of $\pi T$. Thus, the thermal contributions are given by
\begin{equation}
    I_{\text{s}\psi} = I_{\text{s}\la} = I_{\text{g}\psi} = I_{\text{g}\la} = 
    -4\,\frac{c g_0^4}{|a_0|^4} \, \b V \, \Omega_{+-} \,,
\end{equation}
where $\Omega_{+-} \equiv \inlinesumint{p}{+} \inlinesumint{q}{-} 
\frac{(p\cdot q)^2}{p^2 q^2} = \frac{4}{3}J_+ J_-$.

Having deduced the values of the diagrams in Figure \ref{fig:twoloop},
and knowing that $c$ is positive,
we can now understand how $\Ntwo$ gauge theory 
equilibrates at low temperature. The free energy density, viewed 
as a functional of $a_0$, is the effective scalar potential after having
integrated out all thermal fluctuations.
Adding the blackbody and (undetermined) higher-order contributions yields
\begin{equation}
\label{FEneutral}
    (F(a_0)/V)_\text{neutral} 
    = \biggl[-\frac{\pi^2}{12} - \frac{\pi^4}{24}\frac{c g_0^4 T^4}{|a_0|^4}
    + O\Bigl(\frac{g_0^6 \, T^4}{|a_0|^4}\Bigr)\biggr]T^4.
\end{equation}
The `neutral' subscript is just a reminder that this is the contribution
from the neutral degrees of freedom described by the low-energy
effective Abelian theory;
the heavy charged degrees of freedom add the Boltzmann suppressed
contribution \eqref{FEcharged}.
A key feature of the result \eqref{FEneutral} is that
the free energy density decreases (becomes more negative)
as one moves toward smaller values of $|a_0|$.

The subleading $-T^8/|a_0|^4$ behavior of the free energy density has a 
three-loop origin in the microscopic description of the theory. Analogous 
behavior is also observed in IIB supergravity calculations for the 
semiclassical region of $SU(2)$ $\Ntwo$ theory \cite{deMello} 
and in the $SU(\Nc)$ $\Nfour$ theory with $\Nc \to \infty$ and strong 't Hooft
coupling \cite{Tseyt}.%
\footnote{
    In Ref.~\cite{Tseyt}, the supergravity interaction potential between a stack of
    $\Nc$ coincident non-extremal D3-branes and a single ``probe'' D3-brane was
    computed. This potential was interpreted as arising from the Wilsonian effective 
    action for the massless modes obtained by integrating out the massive modes of
    $\Nfour$ theory on its Coulomb branch.
    The potential was found to be attractive. This implies that the
    leading dependence of the free energy on the scalar expectation value also 
    comes with a minus sign. Indeed, Eq. (3.6) of Ref.~\cite{Tseyt} shows this 
    explicitly. It was also argued that the weak-coupling expansion for the 
    free energy density contains a nontrivial $T^8/|a_0|^4$ term.
}
It is also worth noting that the $F^4$ interaction perturbs the free 
Hamiltonian by
\begin{equation}
    -\frac{c \, g_0^4}{64|a_0|^4} \int d^3x\>
    (F_{\mu\nu}+\widetilde{F}_{\mu\nu})^2
    (F_{\rho\s}-\widetilde{F}_{\rho\s})^2
\end{equation}
which is negative semi-definite since $c$ is positive. 
Therefore, the $F^4$ terms lower the classical energy.
[This does not mean that the spectrum is unbounded below ---
the signs of higher powers of $F^2$
become important for large field strengths.]
Reassuringly, similar behavior 
is found in other effective theories
with Abelian gauge fields ({\it e.g.}, the Born-Infeld action for a $U(1)$ 
gauge field localized to a D-brane, or the Euler-Heisenberg action for QED) 
\cite{Adams}.

It is also reassuring to note that
the sum of the free energy density contributions 
at low temperature from charged and neutral fields is consistent
with the high temperature result. More
precisely, the expressions for $F/V$ far out on the free energy surface 
at $T \ll M_W$, and at $T \gg M_W$, match to 
leading order in the coupling. This follows from applying the asymptotic 
formula $h(\rho) \sim -\frac{4}{\pi^2}(2\rho)^{3/2}e^{-\pi\rho}$ (derived in
Ref.~\cite{YamYaf}) to Eq.~\eqref{FEhigh2},
and comparing that to the sum of Eqs.~\eqref{FEcharged} and \eqref{FEneutral}.

Lastly, let us make explicit why instanton contributions to the prepotential
may be ignored compared to the non-holomorphic function $\K$.
Far out on moduli space,
one-instanton corrections to $\F(a)$ take the form $f_1\, a^2(\La/a)^4$ with
$f_1 \neq 0$ \cite{SW1}.
The off-shell form of $\L_\text{eff}^{n=2}$ 
given in expression \eqref{Leff2c} involves only the real and imaginary parts
of $\F''(a)$ and its derivatives.
After inserting Eq.~\eqref{expansion}, suitably 
redefining the coupling constant to include one-instanton effects, and
expanding in powers of $\atilde/a_0$, the terms containing $f_1$ and at least 
one field $\atilde$ begin at $O(\La^4/a_0^5)$. This leads to a subleading 
power correction relative to the leading result \eqref{FEneutral}.

\subsection{Near the massless monopole singularity}

We now switch attention to
the strongly-coupled region of moduli space. 
Vacua near $u = u_0$ have spectra that include two types of BPS states:
electrically charged $W$ bosons with mass $M_W/\La \sim O(1)$ and magnetically 
charged monopoles with mass $M_\text{m}/\La \sim O((u/u_0)-1)$.
We assume the temperature is far below the strong scale,  $T \ll \La$,
but this leaves the 
freedom to consider two distinct regimes: 
(\textit{i}) $T \gg M_\text{m}$ (hot monopoles), or
(\textit{ii}) $T \ll M_\text{m}$ (cold monopoles).

In terms of the monopole dynamics,
case (\textit{i}) is a high temperature regime,
so it is natural to 
construct a three-dimensional effective theory as in
Appendix \ref{app:hightemp}. 
Recall that the low energy theory near $u = u_0$ is an Abelian gauge theory of
$\A_D = (A_D, W_{D\a})$ with hypermultiplet matter $\H = (Q,Q')$ 
of mass $M_\text{m} = \sqrt{2}\,|a_D|$. The monopole
couples locally to the dual photon so this is simply an $\Ntwo$ generalization 
of QED in four dimensions. The effective theory is infrared free with a 
coupling $g_D = g_D(M_\text{m}) \ll 1$. In $\None$ superspace, 
\begin{equation}
\begin{split}
    -g_D^2 \, \L_\text{QED} & =
    \biggl(\int d^2\th\> \fourth W_D^\a W_{D\a} + \text{H.c.}\biggr)
    + \int d^2\th\, d^2\thbar\> A_D^\dag A_D \\
    &
    + \int d^2\th \, d^2\thbar\> \bigl(
    Q^\dag e^{2V_D} Q + Q'^\dag e^{-2V_D} Q' \bigr)
    + \biggl(-i\sqrt{2}\int d^2\th\> Q' A_D Q + 
    \text{H.c.}\biggr).
\end{split}
\end{equation}
The chiral multiplets $Q$ and $Q'$ are oppositely charged under the 
magnetic $U(1)$ gauge group, and under an ordinary $U(1)_\text{f}$ 
flavor symmetry. The superpotential is uniquely fixed by $\Ntwo$ supersymmetry. 
Under $U(1)_R$ transformations, $A_D$ has charge 2, $W_{D_\a}$ has charge 1, 
and $Q$ and $Q'$ are neutral. It follows that both Weyl fermions from the
hypermultiplet have {\it R}-charge $-1$, so conservation of the 
$U(1)_R$-current is anomalous at the one-loop level. 
One may determine the residual symmetry from the fact that 
$\tau_D \sim -\frac{i}{\pi}\ln(a_D)$ must be $2\pi$-periodic under shifts of 
the effective theta angle. The global symmetry is 
$SU(2)_R \times (\Z_4)_R \times U(1)_\text{f}$. 
In components,
\begin{equation}
\label{QED}
\begin{split}
    g_D^2\, \L_\text{QED} &= 
    \fourth F_D^{\mu\nu}F_D^{\mu\nu} 
    + i\labar_D\sbar_\text{E}^\mu\p_\mu\la_D 
    + i\psibar_D\sbar_\text{E}^\mu\p_\mu\psi_D 
    + |\p_\mu a_D|^2 \\
    &\quad 
    + |D^+_\mu q|^2 + |D^-_\mu q'|^2
    + i\psibar_q \sbar_\text{E}^\mu D_\mu^+ \psi_q
    + i\psibar_{q'} \sbar_\text{E}^\mu D_\mu^- \psi_{q'} \\
    &\quad
    + \Bigl[
    i\sqrt{2}\bigl(q\labar_D\psibar_q + q'^*\la_D\psi_{q'} 
    - q\psi_D\psi_{q'} - q'\psi_D\psi_q - a_D \psi_q \psi_{q'}\bigr)
    + \text{H.c.}\Bigr] \\
    &\quad
    + 2|a_D|^2(|q|^2 + |q'|^2) + \half\bigl(|q|^2 + |q'|^2\bigr)^2,
\end{split}
\end{equation}
where $F_D^{\mu\nu} = \p^\mu A_D^\nu - \p^\nu A_D^\mu$ and 
$D_\mu^\pm = \p_\mu \pm iA_{D\mu}$.
The mass term $M_\text{m}$ for the 
hypermultiplet components appears when $a_D$ attains a translationally 
invariant expectation value.

This $\Ntwo$ QED theory is valid below the momentum scale $\La$. The next 
most relevant energy scale is the temperature $T$.
Integrating out thermal fluctuations produces a 
three-dimensional effective theory which we denote as ``QED$_3$.'' 
By construction it will reproduce gauge invariant correlators for distances 
large compared to $T^{-1}$. It is given by
\begin{equation}
    Z = \int 
    \D A_D^i\, \D A_D^0\, \D a_D\, \D q\, \D q'\;
    \exp\Bigl[-\frac{1}{g_{D,3}^2}\int_V d^3x\, \L_{\text{QED}_3}\Bigr]\,,
\end{equation}
with
\begin{equation}
\begin{split}
    \L_{\text{QED}_3} &= f + 
    \fourth (F_D^{ij})^2 + \half(\p_i A_D^0)^2 
    + \half m_\text{E}^2 (A_D^0)^2 + |\p_i a_D|^2 
    + m_\text{s}^2|a_D|^2 \\
    &\quad
    + |D_i^+ q|^2 + |D_i^- q'|^2
    + \bigl(m_\text{h}^2 + (A_D^0)^2 + 2|a_D|^2\bigr)
    \bigl(|q|^2 + |q'|^2\bigr)
    + \half\bigl(|q|^2 + |q'|^2\bigr)^2 \\
    &\quad
    +\d U_\text{thermal}(F_D^{ij}, A_D^0, a_D, q, q').
\end{split}
\end{equation}
The construction is similar in spirit to that for ESYM discussed in Appendix
\ref{app:hightemp}. The fields are all mass
dimension one bosonic zero-frequency modes that have been rescaled so that 
their kinetic terms have canonical 
normalization. To leading order in the dual coupling $g_D$,
$g_{D,3}^2 = g_D^2 T$.
The covariant derivative acting on charged scalar components 
(originally from the hypermultiplet) is 
$D^\pm_i = \p_i \pm i A_{Di}$
and the field strength is $F_D^{ij} = \p^i A_D^j - \p^j A_D^i$.
The effective theory
QED$_3$ has $U(1)$ gauge invariance, plus translation 
and rotation symmetry. The global symmetries are realized as follows: 
$\bigl(\begin{smallmatrix} q \\ q'^*\end{smallmatrix}\bigr)$
transforms as a doublet of $SU(2)_R$, $a_D \to -a_D$ under the
$(\Z_4)_R$ generator,
and $q \to e^{i\w}q$ and $q' \to e^{-i\w}q'$ under 
$U(1)_\text{f}$ for arbitrary real $\w$.
Gauge invariance allows mass terms for the various 
three-dimensional scalar fields. {\it R}-symmetry requires that the 
hypermultiplet scalars appear in the invariant combination $|q|^2 + |q'|^2$,
and that there are no operators cubic in $a_D$. All other local, gauge 
invariant operators of mass dimension 4 or higher are lumped into 
$\d U_\text{thermal}$.

The electrostatic mass $m_\text{E}$, hypermultiplet mass $m_\text{h}$, and 
dual scalar mass $m_\text{s}$ are fixed via matching calculations. One finds
\begin{equation}
    m_\text{E}^2 = m_\text{h}^2 = 2m_\text{s}^2 = g_D^2 T^2 + O(g_D^4 T^2)\,.
\end{equation}
Terms in
$\d U_\text{thermal}$ may be calculated using background field methods.
Note that the tree-level scalar potential in the QED Lagrange density, given by
the last line of Eq.~\eqref{QED}, vanishes when $q = q' = 0$, regardless of the 
value of $a_D$. So in the four-dimensional action one may expand 
around a saddle point $a_D = a_{D0}$ (constant) and all other fields 
zero. Integrating out Gaussian fluctuations around this background leads to the
following effective potential from non-static modes%
\footnote{
    The mean-field-dependent quadratic forms in the shifted Lagrange 
    density involve only the hypermultiplet component fields, so one does not
    need to fix a gauge at leading order.
}
\begin{multline}
    (T/g_{D,3}^2) U_\text{thermal}(a_{D0})\Bigr\vert_\text{all other 
    fields zero} \\ 
    = -\frac{\pi^2}{6}T^4 + \frac{\pi^2}{4}
    \biggl[\frac{M_\text{m}^2}{\pi^2 T^2} 
    + \ln 2\biggl(\frac{M_\text{m}^2}{\pi^2 T^2}\biggr)^2+
    \sum_{n=3}^\infty c_n 
    \biggl(\frac{M_\text{m}^2}{\pi^2 T^2}\biggr)^n\biggr]T^4
    + O(g_D^2T^4) \,,
\end{multline}
where the effective monopole mass
\begin{equation}
    M_\text{m}^2 = 2|a_{D0}|^2.
\end{equation}
An expression for the coefficients $c_n$ is given in Appendix
\ref{app:hightemp}.
In $U_\text{thermal}$, the constant term represents the blackbody radiation
from an Abelian vector and hypermultiplet, and the coefficient of the term 
quadratic in $a_{D0}$ agrees with $m_\text{s}^2$; everything else
constitutes $\d U_\text{thermal}$.

Consider the momentum hierarchy $T \gg M_\text{m} \gg g_D T$.%
\footnote
    {
    The other regime, $M_\text{m} \ll g_D T$, is unremarkable since, 
    for sufficiently small $a_D$, the $O(g_D T)$ screening mass provides
    a big curvature at the origin of field space.
    }
To integrate out
massive monopole fields in QED$_3$, expand the dual scalar field around its
expectation value as $a_D = \vev{a_D} + \s$ with $\vev{a_D} = a_{D0}$. Then
\begin{equation}
\begin{split}
    \L_{\text{QED}_3} &= 
    a_D\, (-\nabla^2 + m_\text{s}^2) \, a_D^* 
    + \half A_D^0 \, (-\nabla^2 + m_\text{E}^2) \, A_D^0 \\
    &\quad + 
    \begin{pmatrix}q^*, & q'\end{pmatrix}
    \begin{pmatrix}-\nabla^2 + m_\text{h}^2 + M_\text{m}^2 & 0 \\
    0 & -\nabla^2 + m_\text{h}^2 + M_\text{m}^2\end{pmatrix}
    \begin{pmatrix}q \\ q'^*\end{pmatrix} + \dotsb \,,
\end{split}
\end{equation}
where the ellipsis indicates terms cubic and higher order in fluctuations. The
static hypermultiplet contribution to the effective potential is 
\begin{equation}
    (T/g_{D,3}^2) U_\text{static} = 4\, I(M_\text{m}^2)T
    \bigl[1 + O(m_\text{h}^2/M_\text{m}^2)\bigr],
\end{equation}
where the function $I$ is given in Eq.~\eqref{loop}.
The overall factor of 4 accounts for the four real
degrees of freedom in $q$ and $q'$.

The new lowest energy effective theory,
valid for distances large compared to $M_\text{m}^{-1}$,
is a three-dimensional $U(1)$ gauge theory
with coupling $g_{D,3}^2$ which also includes a neutral real scalar $A_D^0$
with mass $m_\text{E}$ and a neutral complex scalar $a_D$ with mass
$m_\text{s}$. The free energy density is obtained from the sum of 
$U_\text{thermal}$ and $U_\text{static}$,
\begin{equation}
\label{FEstrong1}
\begin{split}
    F(a_{D0})/V &= T^4\biggl\{
    -\frac{\pi^2}{6} + \frac{\pi^2}{4}\biggl[\frac{M_\text{m}^2}{\pi^2 T^2} 
    + \ln 2\biggl(\frac{M_\text{m}^2}{\pi^2 T^2}\biggr)^2+
    \sum_{n=3}^\infty c_n 
    \biggl(\frac{M_\text{m}^2}{\pi^2 T^2}\biggr)^n\biggr] + O(g_D^2)\biggr\}
\\
    &\quad 
    + M_\text{m}^3T\biggl[-\frac{1}{3\pi} + O(g_D^2T^2/M_\text{m}^2)\biggr]
\\
    &\quad
    + O((g_D T)^3T)\,.
\end{split}
\end{equation}
Each line in Eq.~\eqref{FEstrong1} displays a contribution from one of the three
momentum scales: $T$, $M_\text{m}$, and $g_D T$ (in that order).
Defining a dimensionless mass ratio,
\begin{equation}
    \rho = \frac{M_\text{m}}{\pi T} \,,
\end{equation}
we have
\begin{equation}
\label{FEstrong2}
    F(a_{D0})/V = 
    \Bigl[-\frac{\pi^2}{12} + \frac{\pi^2}{4}\,h(\rho) + O(g_D^2)\Bigr]T^4,
\end{equation}
where $h(\rho)$ is a function that increases monotonically for all $\rho$
and is given explicitly by Eq.~\eqref{eq:h}.
In the interval $1 \gg \rho \gg g_D$, where Eq.~\eqref{FEstrong2} is valid, 
the free energy density is minimized as the monopole mass approaches zero.%
\footnote{
    Ref.~\cite{Wirstam} claims that a nontrivial minimum of the free energy
    density exists at $M_\text{m} \sim O(g_D^2 T)$. Since $M_\text{m} \propto
    |u-u_0|$, this would imply an entire circle of minima in the $u$-plane
    centered around the massless monopole point.
    A continuous set of degenerate 
    equilibrium states suggests a spontaneously broken continuous symmetry.
    What is it? There are only two possibilities: $U(1) \subset SU(2)_R$ or
    $U(1)_\text{f}$, but $a_D$ is not charged under either group. The 
    resolution to this puzzle lies in properly handling the contributions of
    bosonic zero frequency modes to the effective scalar potential. The 
    assertion of Ref.~\cite{Wirstam} follows from balancing the 
    positive $O(M_\text{m}^2 T^2)$ term in Eq.~\eqref{FEstrong1} against the 
    negative $O(g_D^2 M_\text{m} T^3)$ term, and this seems to imply that a 
    minimum occurs at $M_\text{m}/T \sim O(g_D^2)$. However, this is
    outside the range of validity of the calculation and is a misuse of the
    renormalization group. Recall that Eq.~\eqref{FEstrong1} 
    is valid in the regime $1 \gg M_\text{m}/T \gg g_D$.
    For $M_\text{m}/T \ll g_D$, 
    the Wilsonian approach requires integrating out fluctuations 
    whose correlation lengths are set by the Debye screening length 
    $(g_D T)^{-1}$, not $M_\text{m}^{-1}$. Therefore, it is impossible to 
    generate effective interactions which are non-analytic in $M_\text{m}^2$.
    In particular, there is no term linear in $M_\text{m}$ in the effective
    potential.
}

Near the monopole singularity, $a_{D0}$ is mapped
back to the gauge invariant coordinate $u$ via the linear relation 
$a_{D0} \approx c_0(u- u_0)$, where $c_0 = i/(2\La)$ may be determined 
from the elliptic curve solution \cite{SW1}. 
Since the free energy density decreases as $u$ approaches $u_0$,
the point $u_0$ at which monopoles become massless must be a local
equilibrium state.
The effective theory at $u = u_0$ is infrared free,
so the free energy density at this particular 
point is simply the blackbody contributions from a massless vector multiplet
and hypermultiplet, up to corrections suppressed by the strong coupling scale,
\begin{equation}
F/V = -\tfrac{\pi^2}{6}\, T^4 \left(1 + O(T/\La)\right) .
\end{equation}
Using the discrete $R$-symmetry,
this formula for $F/V$ must also hold at $u = -u_0$, the point where dyons 
become massless. Hence, there are two degenerate local minima of the
free energy surface.

Finally, let us consider case (\textit{ii}),
where $T \ll M_\text{m} \ll \Lambda$, so the monopoles are cold and heavy 
and must be integrated out before 
considering the effects of thermal fluctuations. The resulting effective
theory is given to next-to-leading order in the derivative expansion by 
a Lagrange density
\begin{equation}
    \L_{D,\,\text{eff}} = \L_{D,\,\text{eff}}^{n=2} 
    + \L_{D,\,\text{eff}}^{n=4} + O(n \geq 6).
\end{equation}
This describes the 
interactions of a massless $U(1)$ $\Ntwo$ vector multiplet 
$\A_D = (A_D, W_{D\a})$ for distances large compared to $M_\text{m}^{-1}$.
As in Sec.~\ref{subsec:infty}, one may expand around a translationally 
invariant background $a_D(x) = a_{D0}$, define the small coupling 
$g_{D0}^2 = 8\pi^2/(\ln|\La/a_{D0}|^2 - 3)$, and compute the free energy 
density perturbatively in $g_{D0}$ and as an expansion in inverse powers of 
$a_{D0}$.
Since the leading logs in $\F(a)$ and $\F_D(a_D)$ only differ in
functional form by a multiplicative factor, it follows that operators from
$\L_{D,\,\text{eff}}^{n=2}$ do not contribute to $F/V$ (aside from the trivial
blackbody terms) until possibly $O(g_{D0}^8 T^8/|a_{D0}|^4)$ \cite{Wirstam}.%
\footnote{
    Corrections to $\F_D$ of the form $\La^2 \sum_{n=1}^\infty
    c_n (a_D/\La)^n$ arise from integrating out infinitely many massive BPS
    states \cite{Lerche}. This can lead to $O(T/\La)$ corrections in the free
    energy. We assume a separation of scales $T \ll M_\text{m} \lll \La$
    so that $T/\La$ is still smaller than $(T/M_\text{m})^4$.
}
The leading correction to $F/V$ comes from $\L_{D,\,\text{eff}}^{n=4}$. 
No new work is needed to find the correction because electric-magnetic 
duality in the form \eqref{SonK} implies that $\K_D$ and $\K$ have identical
formulas. Hence, we simply adapt the result from Eq.~\eqref{FEneutral}.
The free energy density functional is
\begin{equation}
    F(a_{D0})/V = 
    \biggl[-\frac{\pi^2}{12} 
    - \frac{\pi^4}{24}\frac{c g_{D0}^4 T^4}{|a_{D0}|^4} +
    O\Bigl(\frac{g_{D0}^6 T^4}{|a_{D0}|^4}\Bigr)\biggr]T^4.
\end{equation}
This expression decreases as one moves toward the massless monopole point,
and crosses over into the form \eqref{FEstrong2} when the monopole
mass drops below $T$.

\section{Mass-deformed \boldmath $SU(2)$ $\Nfour$ gauge theory}
\label{sec:star}

A simple generalization of pure $\Ntwo$ gauge theory is the addition of a single
massive elementary hypermultiplet in the adjoint representation.
This theory (often referred to as $\Ntwo^*$) 
is controlled in the far UV by a fixed point (the conformal $\Nfour$ 
gauge theory), but the relevant mass deformation induces running in the 
coupling, so that in the deep IR the theory is again pure $\Ntwo$ gauge theory.
The Lagrange density can be obtained by adding to Eq.~\eqref{Ls} 
the K\"ahler term for the hypermultiplet and the superpotential
\begin{equation}
    W = -i\frac{2}{g^2}\tr\bigl(\sqrt{2}\Phi[Q,Q'] + mQQ'\bigr).
\end{equation}
By a field redefinition, $m$ may be chosen real. One is
free to specify the value of an exactly marginal coupling in the UV. Let 
$q_0 = e^{2\pi i\tau_0}$, where $\tau_0 = \th_0/(2\pi) + i 4\pi/g_0^2$ is any
complex number in the upper half plane. A choice of $q_0$ defines a 
scale $\La_0 \sim |q_0^{1/4}m|$ where the theory evolves into the 
strongly-coupled pure $\Ntwo$ theory \cite{SW2}. We consider the limit of 
weak coupling, $|q_0| \ll 1$, so that a large hierarchy exists between $m$ and 
$\La_0$. 

Classically, moduli space is given by $Q = Q' = 0$ and $[\Phi,\Phi^\dag] = 0$,
so once again, vacua may be described as points in the $u$-plane. Far from the 
origin, at $|u| \gg \La_0^2$, the weak coupling permits a mean field 
analysis. After applying the Higgs mechanism (for $\phi = a\,\s^3/2$), 
one may read off 
the spectrum from the hypermultiplet F-term contributions to the scalar 
potential and the Higgs kinetic term. In $\Ntwo$ language, the vector multiplet
splits into a massless photon $\A^3$ plus charged $W$ bosons $\A^\pm$ with 
masses $\sqrt{2}|a|$. The hypermultiplet splits into a neutral component $\H^3$
with mass $m$ and charged components $\H^\pm$ with masses $|m \mp \sqrt{2}a|$.
A novel feature of this spectrum is that one of the electrically 
charged hypermultiplets (call it an ``electron'')
can become massless at $a = \pm m/\sqrt{2}$. Therefore, in addition to the
singularities where either a magnetic monopole or dyon goes massless, 
there is an additional singularity where an electron becomes massless \cite{SW2}. 
This third singularity is located at $u \approx \frac 14 m^2$.
A simple QED-like effective
theory can be constructed near this point valid on distances $\gg m^{-1}$. By
matching the $SU(2)$ gauge coupling onto the one-loop QED coupling at the scale
$M_W \sim m$, then running the QED coupling down to the mass scale of the light
electron, one obtains the prepotential for an effective Abelian theory 
\cite{BPP},
\begin{equation}
\label{prepotstar}
    \F(a) \sim \tfrac{1}{2} \tau_0 \, a^2 
    + \frac{i}{4\pi}\,a^2\ln\biggl(\frac{a^2}{\La_0^2}\biggr) 
    - \frac{i}{4\pi}\,(a-m/\sqrt{2})^2\ln\biggl(\frac{(a-m/\sqrt{2})^2}{\La_0^2}
    \biggr).
\end{equation}

The perturbative analysis of Sec.~\ref{subsec:infty} may be repeated for the
prepotential \eqref{prepotstar} by expanding around a point near the third 
singularity, $a_0 = m/\sqrt{2} + \Delta a_0$. A similar 
conclusion is reached: the prepotential cannot contribute to the free energy
density until at least $O(\bar{g}_0^8 \, T^8/|\Delta a_0|^4)$, where 
$1/\bar{g}_0^2 = 1/g_0^2 + \frac{1}{4\pi^2}
\ln\bigl|\frac{m/\sqrt{2}}{\Delta a_0}\bigr|$. 

Instead of studying in detail four derivative terms in the effective action 
determined by a non-holomorphic function $\K$, one may argue that the third 
singularity must be a local minimum of the effective potential as follows. 
Since the electron becomes massless at $u \approx \frac 14 m^2$, the effective 
coupling $\bar{g}_0$ vanishes at this point and the low energy theory is free. 
For low temperatures, $T \ll \La_0$, the free energy density at the third
singularity is simply $F/V = -\frac{\pi^2}{6}\,T^4(1 + O(T/\La_0, T/m))$.%
\footnote{
    Recall that free $\Ntwo$ vector and hypermultiplets each contribute 
    $-\frac{\pi^2}{12}\,T^4$ to the free energy density.
}
Consider turning
off the temperature and choosing a vacuum close to the third singularity. 
The spectrum still includes a massless photon, but the electron will have some 
small non-zero mass $m_\text{e}$.
Now turn on a temperature $T \ll m_\text{e}$. It is 
difficult to thermally excite electrons so their contribution to $F/V$ will be
exponentially suppressed. Since the effective theory near this point is 
weakly-coupled, we may trust the blackbody approximation to the free energy, 
but now this comes from a {\it single} type of $\Ntwo$ multiplet rather than 
two. Hence, the third singularity must lie deeper in the free energy surface 
than nearby points.

Finally, we must understand the behavior of the free energy density for 
$|u| \gg m^2$. One cannot simply adapt the $\Ntwo$ result since $\Ntwo^*$ is 
UV conformal, rather than asymptotically free. The non-holomorphic
function $\K$ for $\Nfour$ theory is believed to be exactly of the form given in
Eq.~\eqref{K} without suffering renormalization \cite{DS,KuzMc2,Kuz}. 
The coefficient is known to be $c = 1/\pi^2$
\cite{Peri}. With positive $c$, thermal fluctuations 
again make the asymptotic region of the free energy surface locally unstable.

\section{Discussion}
\label{sec:discuss}

The non-trivial moduli space of
$SU(2)$ $\Ntwo$ supersymmetric Yang-Mills theory leads to
a rich variety of dynamics on multiple length scales.
Analyzing the thermodynamics of the theory requires
careful application of effective field theory techniques
to disentangle contributions from different types of fluctuations.
At high temperature, 
$T \gg \La$, we found a unique $\Z_2$-invariant equilibrium state 
with a free energy density given by Eq.~\eqref{FEhigh3}. 
At low temperatures,
the flat zero-temperature ground state
energy surface deforms into a non-trivial free energy surface.
Far from the origin of moduli space, where $M_W \gg \La$,
we found that an arbitrarily small temperature, $T \ll M_W$,
causes the free energy surface to rise asymptotically.
This corrects previously reported results \cite{Wirstam}
on the thermodynamics of this theory,
and implies that minima of the free energy surface must lie
in the portion of moduli space where the gauge coupling is strong.
By using the dual description of the theory near the massless monopole
(or dyon) points, we were able to analyze the thermodynamics when
$M_{\rm m} \ll T \ll \La$ and monopoles (or dyons) are hot,
as well as when $T \ll M_{\rm m} \ll \La$ and monopoles (or dyons) are cold.
We found that the free energy surface has degenerate local minima
at the massless monopole and dyon points.
Our results are summarized in Figure \ref{fig:uflow}(b).

As there are no points in moduli space with enhanced gauge symmetry,
the simplest scenario consistent with the above observations is to assume
that there are no other local minima of the free energy density,
and that the free energy surface smoothly decreases toward the
massless monopole and dyon points throughout the intermediate regions
in which no weak coupling description is applicable.
This gives a simple, consistent picture in which 
the discrete $R$-symmetry is spontaneously broken
at low temperatures, with two co-existing equilibrium states.

The restoration of $R$-symmetry at high temperature,
combined with its spontaneous breakdown at low temperature,
implies that there must be a genuine thermodynamic phase transition.
The transition temperature must be some pure number times $\La$,
which is the only intrinsic scale in the theory.
The spontaneous symmetry breaking, and consequent change in the
number of distinct equilibrium states,
ultimately arises from the existence of multiple special points in moduli 
space where equal numbers of massless states appear in the low energy spectrum. 
It is instructive to contrast this with $SU(\Nc)$ $\Nfour$ gauge theory.
The moduli space of this theory is locally flat and corresponds to the 
orbifold $\R^{6(\Nc-1)}/S_{\Nc}$. Only the single vacuum state at the origin
is a fixed point of the entire permutation group.
At this point the theory is superconformal and has the maximal number
of massless gluons in its low energy spectrum.
At weak coupling, these gluons provide the largest possible order one
contribution to the free energy density in the form of blackbody radiation.
At any non-zero temperature, there is a unique equilibrium state.

We also examined weakly-coupled $SU(2)$ $\Ntwo^*$ theory.
At low temperature,
we found that the additional hypermultiplet leads to the appearance
of a third local minima in the free energy surface.
This suggests the possibility of three distinct 
thermal equilibrium states.
They correspond to points in the $u$-plane where a
hypermultiplet (either solitonic or elementary) becomes massless.
All three local minima on the free energy surface
have the same free energy density,
$F/V = -\frac{\pi^2}{6} \, T^4$,
up to corrections suppressed by the ratio of temperature to the
strong coupling scale $\La_0$.
As only two local minima are related by the discrete
{\it R}-symmetry, we are unable to determine,
based on our effective theory analysis,
whether the three local minima are exactly degenerate,
and if not which are the true global minima.

An obvious extension of this work would be to generalize the low temperature
analysis to $SU(\Nc)$ gauge groups with any number of colors.
In particular,
it would be interesting to study thermodynamics in the weakly-coupled 
$SU(\Nc)$ $\Ntwo^*$ theory.
If an $R$-symmetry phase transition 
exists in this theory, one could parametrize the transition temperature as 
$T_c = m f(\Nc, \La_0/m)$ for some dimensionless function $f$. 
An understanding of the large $\Nc$ limit of this formula might shed light on 
recent results obtained from the supergravity dual for the strongly-coupled 
version of the theory at finite temperature \cite{BDKL}. 
In the strong-coupling limit, the scale $\La_0$ is the same as the mass 
deformation $m$. Therefore, the critical temperature of the transition may be 
parametrized as $T_c = m \tilde{f}(\Nc)$ for some unknown function
$\tilde{f}$.
A phase transition can be detected, in the large $\Nc$ limit, by
finding a zero of the free energy density as a function of $m/T$. However, 
numerical calculations do not show any zero-crossing behavior in the interval
$0 \leq m/T \lesssim 12$ \cite{Buch}. 
This is somewhat unexpected, as one might have
expected qualitatively different behavior
in the regimes $T \ll m$ and $T \gg m$ \cite{BL}.
One possible resolution would be for the function 
$\tilde{f}(\Nc)$ to scale as some (positive) power of $1/\Nc$,
forcing $m/T_c$ to move off to infinity as $\Nc \to\infty$.
Our work was originally motivated by the desire to 
better understand the nature of phase transitions in non-conformal gauge 
theories with supergravity duals. One of the basic questions for the 
thermodynamics of such theories is understanding
how the large $\Nc$ limit affects the 
number and location of thermal equilibrium states.

\acknowledgments

S.P. is indebted to Andreas Karch, Ann Nelson, M\aa ns Henningson, and 
Sergei Kuzenko for helpful conversations and correspondence. We thank the
JHEP referee for pointing out how the non-holomorphic function $\K$ in 
$\N=2$ theory may be
obtained from existing superfield calculations in $\Nfour$ theory.
Comments on the manuscript from Ethan Thompson, Carlos Hoyos, and Alex
Buchel are much appreciated. 
This work was supported in part by the U.S. Department of 
Energy under grant DE-FG02-96ER40956.

\appendix

\section{High temperature effective theory}
\label{app:hightemp}

In this appendix we compute the thermal effective potential for the complex 
scalar field $\phi$ along its flat directions in $\Ntwo$ gauge theory. 
A unique global minimum is shown to be located at vanishing $\phi$.
Our procedure closely follows the treatment for weakly-coupled 
$\Nfour$ gauge theory in Ref.~\cite{YamYaf}.

The most important effect of thermal fluctuations 
will be to generate temperature-dependent effective masses for certain
fields.
These are the complex scalar field $\phi$ and the ``time'' component $A_0$ of
the real Euclidean gauge field.%
\footnote{
    More precisely, one may view $A_0$ as the traceless part of the
    gauge invariant Polyakov loop around the thermal circle \cite{ArnYaf}.
}
At high temperature the effective coupling $g(T)$ is small,
so these masses are
calculable in perturbation theory, are positive, and $O(gT)$.%
\footnote{
    Masses computed in strict perturbation theory should be understood 
    as matching parameters in the low energy effective theory.
    The physical Debye screening mass, for example,
    includes nonperturbative corrections of order $g^2T$ \cite{ArnYaf}.
}
Consequently, there is positive curvature in field configuration space at
$\phi = A_0 = 0$.
[This potential for $A_0$ is consistent with the expected 
spontaneous breaking of center symmetry at high temperature,
since a vanishing mean and small fluctuations for $A_0$ imply
a non-zero mean for
the Polyakov loop, $\tr \, \mathcal {P} (e^{i \int\! A_0 dt})$.]
The point $\phi = 0$ is a local minimum of the effective scalar 
potential and therefore also of the free energy surface. The goal is to now 
sharpen this observation and determine whether this local minimum
is actually a global minimum of the effective scalar potential.

Consider a regime where the temperature is much greater than the mass of 
$W$ bosons, $T \gg M_W$. 
Finite temperature effects on the long distance properties of 
the theory are captured by constructing a sequence of effective field theories.
For an asymptotically free non-Abelian gauge theory at high temperature, 
the momentum scales shown in Eq.~\eqref{hierarchy} are important for understanding
its static properties. The temperature $T$ is the typical momentum of particles
in the plasma, the Debye scale $gT$ sets the (inverse) correlation length of
color-electric screening, and the magnetic mass gap $g^2T$ sets the (inverse)
correlation length of color-magnetic screening. The separation between these 
scales is parametrically large at weak coupling. 
One may construct a low energy effective theory known as ``ESYM'' that 
reproduces static gauge invariant correlators of $SU(2)$ $\Ntwo$ gauge 
theory for distances much greater than the inverse temperature \cite{YamYaf}.

The starting point for the construction is to consider the 
theory defined in a periodic spatial box of volume $V$ and inverse temperature
$\b$. The partition function has the functional integral 
representation
\begin{equation}
    Z = \int \D A_\mu\, \D\la\, \D\psi\, \D\phi\; 
    \exp\biggl[-\frac{1}{g^2} \int_0^\b dx^0_\text{E} \int_V d^3x\, \L\biggr],
\end{equation}
where $\L$ is given by the right hand side of Eq.~\eqref{Lc}.
The Euclidean time direction is compact so the Hilbert space trace 
which defines the partition function requires periodic 
(antiperiodic) boundary conditions for bosonic (fermionic) fields. 
Each four-dimensional bosonic (fermionic) field may be decomposed 
as discrete Fourier series with frequencies that are even (odd) 
integer multiples of $\pi T$ and coefficient functions that depend only on the
three spatial directions. Integrating out all of the non-static modes 
({\it i.e.}, modes with frequencies $|\nu| \geq \pi T$) results in
\begin{equation}
    Z = \int \D A_i\, \D A_0\, \D \phi \;
    \exp\Big[-\frac{1}{g_3^2}\int_V d^3x\, \L_\text{ESYM}\Bigr] \,,
\end{equation}
with
\begin{equation}
\label{ESYM}
\begin{split}
    \L_\text{ESYM} &= f 
    + 2\, \tr\, \biggl\{\fourth F_{ij}^2 
    + \half(D_i A_0)^2 + \half m_\text{E}^2 A_0^2
    + |D_i\phi|^2 + m_\phi^2|\phi|^2
    + \bigl|[A_0,\phi]\bigr|^2 + \half[\phi,\phi^\dag]^2
    \biggr\} \\
    & \quad + \d U_\text{thermal}(F_{ij},A_0,\phi)\,.
\end{split}
\end{equation}
This effective theory
is a three-dimensional theory with $SU(2)$ gauge invariance,
translation and rotation symmetry, and $\Z_8$ {\it R}-symmetry 
which is realized as $\phi \to e^{i\pi/2}\phi$.%
\footnote{
    $SU(2)_R$ invariance holds trivially since the fermion doublet 
    has been integrated out, leaving only $SU(2)_R$ singlets. 
}
The Lagrange density \eqref{ESYM} is obtained from dimensional reduction;
this automatically produces the familiar gauge invariant derivative terms and 
commutator potentials. The adjoint covariant derivative 
$D_i = \p_i + i[A_i,\cdot]$.
%% and the field strength
%% $F_{ij} \equiv \p_i A_j - \p_j A_i + i[A_i, A_j]$.
Gauge invariance allows mass terms for the 
three-dimensional scalar fields $A_0$ and $\phi$ --- these terms 
are non-zero as a result of integrating out the infinite tower of 
non-static modes. The discrete {\it R}-symmetry does not permit terms cubic 
in $\phi$. All other local, gauge invariant operators 
of mass dimension 4 and higher that can be constructed out of $F_{ij}$, 
$A_0$, and $\phi$ are contained in $\d U_\text{thermal}$. The only
operator with dimension zero is the identity and we include it explicitly with
coefficient~$f$.

The three-dimensional fields of ESYM have engineering dimension one and 
represent renormalized zero-frequency modes of the corresponding 
four-dimensional fields.%
\footnote{
    For example, the zero-frequency mode of the field $A_0(x)$ is given by 
    $A_0^{(0)}(\vec{x}) = \b^{-1}\int_0^\b dx^0_\text{E}\, A_0(x)$. 
    The renormalized electrostatic field is defined by 
    $A^\text{ren}_0(\vec{x}) \equiv Z_{A_0}^{1/2} A_0^{(0)}(\vec{x})$, where
    $Z_{A_0}$ is a dimensionless function of the coupling and $\mu/T$, 
    with $\mu$ an arbitrary renormalization scale.
    The wavefunction renormalization factor $Z_{A_0}$
    may be computed in perturbation theory and
    equals $1 + O(g^2 \ln(\mu/T))$.
    To the order at which we will be working
    this finite renormalization factor, as well as those of the other
    fields in the $3d$ effective theory, may be ignored.
    %This wavefunction renormalization is important only if one is trying to get
    %the next-to-leading order correction to the effective potential. The
    %$O(g^0)$ part of the effective potential is the same whether expressed
    %in terms of unrenormalized or renormalized fields.
}
One may choose their normalization so that
$\L_\text{ESYM}$ contains canonically normalized kinetic terms.
To avoid unnecessary clutter in the presentation, we have
chosen not to introduce separate notation for the renormalized
fields of the effective theory.
This slight notational sloppiness is harmless because the 
computation and results for the effective scalar potential presented in this 
appendix are unambiguous to leading order in the coupling.

The coefficients of the operators in the effective theory are determined by
computing correlators of static observables as a function of their spatial 
separation, at short distance,
in the original $\Ntwo$ gauge theory and in ESYM.
Perturbative calculations may be used to match
results within the window%
\footnote{
    This range of scales is determined from the following considerations. Naive
    perturbative calculations in the short distance theory do not account for
    screening of color-electric fields in the plasma which cause $A_0$ 
    to develop a finite correlation length of order $(gT)^{-1}$. Far below this
    length scale, $A_0$ behaves like a massless field and strict 
    perturbation theory can be used \cite{BN2}. The long distance theory 
    includes the effects of integrating out thermal fluctuations with 
    frequencies $\geq \pi T$. Therefore, the effective theory is only valid 
    for energies far below this scale.
}
$(\pi T)^{-1} \ll |\vec{x}{-}\vec{y}| \ll (gT)^{-1}$.
The resulting ESYM effective theory, with all coefficients determined,
then provides a description of physics
valid for distances large compared to $T^{-1}$.

The inverse $3d$ gauge coupling, $1/g_3^2 \equiv Z_g^2\, \b/g^2$,
is an overall factor multiplying the action of the $3d$ effective theory.
We will only need to know $g_3^2$ to leading order,
in which case
the renormalization factor $Z_g^2$ may be replaced by unity.
Hence $g_3^2 = g^2 T + O(g^4 T)$.
Following the techniques described in the appendices of Ref.~\cite{YamYaf},
one finds that the coefficient of the identity operator,
which represents the contribution to the free energy density from the 
momentum scales of order $T$ and up
[according to $F/V = (T/g_3^2)f + \dotsb$],
is given by
\begin{equation}
\label{fESYM}
    f = 3g_3^2T^3\biggl[-\frac{\pi^2}{12} + \frac{g^2}{8} + O(g^4)\biggr].
\end{equation}
The leading term is just blackbody radiation while the $O(g^2)$ term
comes from two-loop bubble diagrams.
By computing the gluon self-energy tensor at zero external momentum in a loop
expansion, one finds the electrostatic mass 
parameter $m_\text{E}$. At leading order it is equal to the Debye mass,
\begin{equation}
\label{staticmass2}
    m_\text{E}^2 = 2g^2T^2 + O(g^4T^2) \,.
\end{equation}
A similar process determines the scalar mass parameter,
\begin{equation}
\label{scalarmass2}
    m_\phi^2 = g^2T^2 + O(g^4T^2)\,.
\end{equation}
The results \eqref{fESYM}--\eqref{scalarmass2} were also 
obtained in Ref.~\cite{Vaz} by studying the thermodynamics
of six-dimensional $\None$ gauge theory. This theory has eight supercharges and
reproduces four-dimensional $\Ntwo$ gauge theory upon dimensional reduction.

Let us pause to understand our earlier claims about thermal masses for 
scalar fields. Like the zero temperature case, there is a commutator potential
of the form $[\phi,\phi^\dag]^2$ that requires the expectation value
of the three-dimensional field $\phi$ to be a diagonal matrix 
(up to gauge transformations). A new feature in the effective 
theory is the presence of the quadratic potential $m_\phi^2|\phi|^2$. 
This term levies an energy cost for
non-zero eigenvalues of $\phi$. For sufficiently 
small $\phi$, this term will completely dominate the behavior of the scalar 
potential. These tree-level considerations indicate that $\phi\,{=}\,0$ 
(along with $A_0\,{=}\,0$) is a local minimum of the free energy density 
function. However, a tree-level analysis is not truly complete without 
understanding the effects of higher dimension operators in 
$\d U_\text{thermal}$. 
In principle one may fix the coefficients of these operators by matching 
successively higher $n$-point functions, but it is more convenient
to use a background field method. 

The idea of the background field method is to choose a general constant
value of the scalar field (other than $\phi = 0$) and evaluate the 
thermal effective potential as a function of $\phi$. We choose
to expand $\phi$ around $\phi_\text{flat} \equiv a\,\s^3/2$,
with the gauge field and fermions vanishing,
thus incurring no energy cost from gradients or tree-level potentials.
Integrating out non-static fluctuations around this background 
field yields all higher order terms in the effective potential when
the scalar field lies along its flat directions,%
\footnote{\label{fn:landau}%
    In the shifted action, there is a bilinear term arising from the 
    covariant derivative of $\phi$ that mixes the fluctuating part of 
    $\phi$ and $A_\mu$. Integrating by parts shows that this term is
    proportional to the divergence of the gauge field. We choose Landau gauge,
    $\p_\mu A^\mu = 0$. Diagrammatically, 
    the bilinear term generates a scalar-vector vertex with an external 
    $\phi_\text{flat}$ leg at zero momentum. The Feynman rule for the vertex 
    is proportional to the momentum of the fluctuating gauge field,
    but a Landau gauge vector propagator projects onto the space
    transverse to the momentum \cite{CW}.
    Thus, with this gauge choice one can ignore the
    bilinear mixing term.
}
\begin{multline}
\label{Uthermal}
    (T/g_3^2)\,
    U_\text{thermal}(\phi_\text{flat}{=}a\s^3/2)\Bigr\vert_\text{all 
    other fields zero}\\ 
    = -\frac{\pi^2}{4}T^4
    +\frac{\pi^2}{2}\biggl[\frac{M_W^2}{\pi^2 T^2} 
    + (\ln 2)\biggl(\frac{M_W^2}{\pi^2 T^2}\biggr)^2+
    \sum_{n=3}^\infty c_n 
    \biggl(\frac{M_W^2}{\pi^2 T^2}\biggr)^n\biggr]T^4
    + O(g^2T^4) \,,
\end{multline}
where $M_W$ denotes the effective $W$ mass,
\begin{equation}
    M_W^2 = 2|a|^2 \,,
\end{equation}
and
\begin{equation}
    c_n = 8(-1)^n\left(1-4^{2-n}\right)
    \frac{(2n{-}5)!!}{(2n)!!}\,\zeta(2n{-}3).
\end{equation}
The coefficients $c_n$ are derived in Appendix C of Ref.~\cite{YamYaf}.
The quantity $(T/g_3^2)\,U_\text{thermal}$ represents the contribution to the
free energy density $F/V$ from the momentum scale $T$ (and above),
and Eq.~\eqref{Uthermal} 
is the result of integrating out only Gaussian 
fluctuations around the background field $\phi_\text{flat}$.
In $U_\text{thermal}$, the constant term represents the blackbody radiation
for a triplet of massless vector multiplets, and the coefficient of the term 
quadratic in $a$ agrees with $m_\phi^2$. 
We define $\d U_\text{thermal}$ to be everything in 
$U_\text{thermal}$ aside from the constant and quadratic terms.

At this point,
we have only assumed that $T \gg M_W$. We now refine the 
restriction to $T \gg M_W \gg gT$. If a nontrivial minimum of the effective 
potential exists, then it must lie in this region. To see this, it is helpful
to define a dimensionless rescaled mass,
\begin{equation}
    \rho \equiv \frac{M_W}{\pi T}\,.
\end{equation}
At lowest order in the coupling, expressed in terms of $\rho$, the 
contribution of non-static modes to the effective
potential takes the form $\frac{\pi^2}{2}f(\rho)\,T^4$,
where $f(\rho) = -\half + \rho^2 + (\ln 2)\, \rho^4 + O(\rho^6)$
is completely independent of $g$.
Including the effects of static modes will be shown
to supplement $f$ with an additional cubic term,
$b\rho^3$. The coefficient $b$ is negative and $O(g^0)$.
Therefore, a nontrivial minimum of the effective potential can
exist only for some $O(g^0)$ value of $\rho$.

To complete the analysis, one must consider the
contribution to the free energy density from fluctuations 
of the static ESYM fields with masses set by $M_W$. 
One may do this by expanding $\phi$ around its expectation value 
and integrating out all fields that develop finite 
correlation lengths of order $M_W^{-1}$. 
Let $\phi = \vev{\phi} + \s$ with 
$\vev{\phi} = a\s^3/2$ and $\s$ a fluctuating adjoint representation complex
scalar field. This field redefinition naturally generates
mass terms that depend on $\vev{\phi}$, as well as new
interactions involving three or more fluctuating fields.
Since the lowest order contributions to the free energy density
are functional determinants obtained by integrating Gaussian fluctuations,
we restrict our discussion to the quadratic forms in $\L_\text{ESYM}$
obtained after the background field shift. For static gauge fields, 
\begin{equation}
\label{qform_mag}
    \L_\text{ESYM} \supset
    \half A_i^3 \Delta_{ij} A_j^3 + W_i (\Delta_{ij} + M_W^2 \d_{ij}) W_j^*,
\end{equation}
where $\Delta_{ij} = -\d_{ij}\nabla^2 + (1-\a^{-1})\p_i\p_j$ is the covariance
operator for vector fields and $W_i \equiv (A_i^1 - i A_i^2)/\sqrt{2}$. 
A Lorentz gauge-fixing term has been included. We choose to compute in Landau 
gauge ({\it i.e.}, we send $\a \to 0^+$).
%\footnote{
%\label{mixedbilinear}%
%    Landau gauge removes the mixed bilinear term 
%    $2\tr\bigl(-i\p_i\s[A_i,\vev{\phi}]^\dag + \text{H.c.}\bigr)$.
%}
Expression \eqref{qform_mag} indicates that an $M_W$-dependent correlation 
length is generated for the off-diagonal components of the static gauge field. 
For the electrostatic scalar, 
\begin{equation}
\label{qform_elec}
    \L_\text{ESYM} \supset
    \half A_0^3 (-\nabla^2 + m_\text{E}^2) A_0^3 
    + W_0 (-\nabla^2 + m_\text{E}^2 + M_W^2) W_0^*,
\end{equation}
where $W_0 \equiv (A_0^1 - i A_0^2)/\sqrt{2}$. Expression \eqref{qform_elec}
shows that off-diagonal components of the electrostatic scalar also obtain 
$M_W$-dependent masses. For the complex scalar,
\begin{equation}
\label{qform_scalar}
    \L_\text{ESYM} \supset
    \phi^3 (-\nabla^2 + m_\phi^2) \, \phi^{3*} + \Sigma^\dag
    \begin{pmatrix}-\nabla^2 + m_\phi^2 + |a|^2 & -a^2 \\
    -(a^*)^2 & -\nabla^2 + m_\phi^2 + |a|^2
    \end{pmatrix}\Sigma,
\end{equation}
where $\Sigma = \frac{1}{\sqrt{2}}
\Bigl(\begin{smallmatrix}\s^1-i\s^2 \\ 
\s^{1*}-i\s^{2*}\end{smallmatrix}\Bigr)$. Since the mass matrix in expression
\eqref{qform_scalar} has eigenvalues $m_\phi^2$ and $m_\phi^2+M_W^2$, there is 
only one (complex) component of $\Sigma$ that obtains an $M_W$-dependent mass. 
Notice that only off-diagonal components of the scalar field 
receive $a$-dependent mass shifts.

Integrating over the off-diagonal fields in
expressions \eqref{qform_mag}--\eqref{qform_scalar} with correlation lengths of 
order $M_W^{-1}$ is straightforward to leading order,
since the integrals are Gaussian.
The basic ingredient needed to evaluate the resulting functional
determinants is the single loop integral
\begin{equation}
\label{loop}
    I(m^2) \equiv \half\int\frac{d^3k}{(2\pi)^3}\>\ln(\vec{k}^{\,2} + m^2) = 
    -\frac{1}{12\pi}\,(m^2)^{3/2}\,.
\end{equation}
(We have regulated the theory by dimensional continuation to
$d = 3{-}2\e$ dimensions.%
\footnote
{
  More precisely, one should first subtract from the integrand a
  term of the form $\ln(\vec{k}^2 + \mu^2)$ where $\mu$ is an IR regulator.
  Alternatively, one may define $I(m^2)$ by
  differentiating the formal integral with respect to $m^2$. The
  resulting integral is only linearly sensitive to the UV cutoff.
  Analytically continuing in dimension and then integrating back
  with respect to $m^2$ leads to the stated result.
})
The contribution to the effective potential from the static modes
with masses of order $M_W$ is simply
\begin{equation}
\label{Ustatic}
    (T/g_3^2)\, U_\text{static} = 2\cdot 4\, I(M_W^2) \, T
    \Bigl[1 + O(g^2T^2/M_W^2)\Bigr].
\end{equation}
The factor of 2 accounts for the complex nature of all the off-diagonal fields.
The factor of 4 counts the total number of propagating fields: 2 for the gauge 
field since there are two transverse directions to a given spatial momentum,
and 2 for the two kinds of scalars.
The relative $O(g^2 T^2/M_W^2)$ corrections in 
Eq.~\eqref{Ustatic} come from evaluating $I$ with the complete mass,
either $m_{\phi}^2 + M_W^2$ or $m_{\text{E}}^2 + M_W^2$,
and then expanding the result in powers of the small ratio
$(m_{\phi}/M_W)^2$ or $(m_{\text{E}}/M_W)^2$.
The minus sign in Eq.~\eqref{loop} is 
physically significant, as it shows that fluctuations of static fields
have a destabilizing effect on the potential. If the magnitude of such 
destabilizing terms are large enough, then a nontrivial minimum will be 
produced away from the origin.

The final low energy effective theory, valid for distances large
compared to $M_W^{-1}$,
is a three-dimensional $U(1)$ gauge theory 
with coupling $g_3^2$, a neutral real scalar $A_0^3$ with 
mass $m_\text{E}$, and a neutral complex scalar $\phi^3$ with mass $m_\phi$.
The effective scalar potential in this theory, for $\phi$ along 
its flat directions, is given by the sum of $U_\text{thermal}$ and 
$U_\text{static}$. It follows that the free energy density, viewed as a 
functional of $a$, is given by 
\begin{equation}
\label{FEhigh1}
\begin{split}
    F(a)/V
    &= 
    T^4\biggl\{-\frac{\pi^2}{4}
    +\frac{\pi^2}{2}\biggl[\frac{M_W^2}{\pi^2 T^2} 
    + \ln 2\biggl(\frac{M_W^2}{\pi^2 T^2}\biggr)^2+
    \sum_{n=3}^\infty c_n 
    \biggl(\frac{M_W^2}{\pi^2 T^2}\biggr)^n\biggr]
    + O(g^2)\biggr\} \\
    & \quad + M_W^3T\biggl[-\frac{2}{3\pi}
	+ O\!\left(\frac{g^2T^2}{M_W^2}\right)\biggr]
\\ & \quad
    + O((gT)^3\,T)\,.
\end{split}
\end{equation}
Each line in expression \eqref{FEhigh1} 
displays the contribution from one of the three
momentum scales: $T$, $M_W$, and $gT$ (in that order).
The contribution from the soft scale
$gT$ may be obtained by integrating out the neutral scalars. 
In terms of the normalized mass $\rho = \sqrt 2 \, |a|/(\pi T)$, 
the final result is
\begin{equation}
\label{FEhigh2}
    F(a)/V = \Bigl[-\frac{\pi^2}{12} + \frac{\pi^2}{2}\,h(\rho) + O(g^2)
    \Bigr]T^4 \,,
\end{equation}
where
\begin{equation}
    h(\rho) \equiv -\tfrac{1}{3}+\rho^2 -\tfrac{4}{3}\,\rho^3+(\ln 2)\,\rho^4
    + \sum_{n\geq 3} c_n\,\rho^{2n}.
\label{eq:h}
\end{equation}
The function $h$ increases monotonically for all $\rho$
(see Ref.~\cite{YamYaf} for a discussion of its global behavior).
Our result for 
$F(a)/V$, which may be trusted in the interval $1 \gg \rho \gg g$, 
is minimized as $\rho$ approaches zero.

In summary, the free energy density is given by
Eqs.~\eqref{FEhigh2}--\eqref{eq:h} for $1 \gg \rho \gg g$.
Since this function is monotonic, the free energy 
density is minimized precisely where the scalar field 
eigenvalues vanish. Hence, $a = 0$ is the unique equilibrium state
in the high temperature regime, $T \gg \Lambda$.

\end{document}